\documentclass[preprintnumbers,article,amsmath,amssymb,floatfix,10pt,prd,twocolumn,superscriptaddress,nofootinbib]{revtex4-2}
\usepackage{bm}
\usepackage{amsfonts}
\usepackage{latexsym}
\usepackage[latin1]{inputenc}
\usepackage{graphicx}
\usepackage{amsmath}
\usepackage{palatino}
\usepackage{mathpazo}
\usepackage{textcomp}
\usepackage{lineno}
\usepackage{float}
\usepackage{booktabs}
\usepackage{dcolumn}
\usepackage{ragged2e}
\usepackage{hyperref}
\usepackage{tensor}
\usepackage[version=3]{mhchem}
\hypersetup{colorlinks,citecolor=blue}
\hypersetup{colorlinks=true,linkcolor=red,filecolor=magenta,    urlcolor=blue}
\usepackage{amsmath}
\usepackage{xcolor}
\usepackage{enumitem}
\usepackage{orcidlink}
\usepackage{epsfig}
\usepackage{subfigure}
\usepackage{commath}
\usepackage{caption}

\def\jnl@style{\it}
\def\aaref@jnl#1{{\jnl@style#1}}

\def\aaref@jnl#1{{\jnl@style#1}}

\def\aj{\aaref@jnl{AJ}}                   
\def\apj{\aaref@jnl{ApJ}}                 
\def\apjl{\aaref@jnl{ApJ}}                
\def\apjs{\aaref@jnl{ApJS}}               
\def\apss{\aaref@jnl{Ap\&SS}}             
\def\aap{\aaref@jnl{A\&A}}                
\def\aapr{\aaref@jnl{A\&A~Rev.}}          
\def\aaps{\aaref@jnl{A\&AS}}              
\def\mnras{\aaref@jnl{Mon.~Not.~Roy.~Astron.~Soc.}}             
\def\prd{\aaref@jnl{Phys.~Rev.~D}}        
\def\prc{\aaref@jnl{Phys.~Rev.~C}}  
\def\prl{\aaref@jnl{Phys.~Rev.~Lett.}}    
\def\qjras{\aaref@jnl{QJRAS}}             
\def\skytel{\aaref@jnl{S\&T}}             
\def\ssr{\aaref@jnl{Space~Sci.~Rev.}}     
\def\zap{\aaref@jnl{ZAp}}                 
\def\nat{\aaref@jnl{Nature}}              
\def\aplett{\aaref@jnl{Astrophys.~Lett.}} 
\def\apspr{\aaref@jnl{Astrophys.~Space~Phys.~Res.}} 
\def\physrep{\aaref@jnl{Phys.~Rep.}}      
\def\physscr{\aaref@jnl{Phys.~Scr}}       
\def\commat{\aaref@jnl{Comm.~Math.~Phys.}}              
\def\science{\aaref@jnl{Science}}               
\def\cqg{\aaref@jnl{Classical Quant.~Grav.}}            
\def\jpcs{\aaref@jnl{JPCS}}                                     
\def\ijmpd{\aaref@jnl{Int.~J.~Mod.~Phys.~D}}                    
\def\grg{\aaref@jnl{Gen.~Relat.~Gravit.}}               
\def\rpp{\aaref@jnl{Rep.~Prog.~Phys.}}          
\def\npa{\aaref@jnl{Nucl.~Phys.~A}}        
\def\lrr{\aaref@jnl{Living Rev.~Rel.}}                   
\def\jcap{\aaref@jnl{J.~Cosmology Astropart.~Phys.}}    
\def\rmp{\aaref@jnl{Rev.~Mod.~Phys.}}   
\def\epjc{\aaref@jnl{Eur.~Phys.~J.~C}} 
\def\plb{\aaref@jnl{~Phy.~Lett.~B}} 
\def\mpla{\aaref@jnl{Mod.~Phy.~Lett.~A}} 
\def\arxiv{\aaref@jnl{arxiv.org}}

\allowdisplaybreaks[1]
\renewcommand{\arraystretch}{1.1}
\begin{document}

\title{Observational constraints on viscous free-$\gamma$ fluid in $f(Q)$ gravity}

\author{Simran Arora\orcidlink{0000-0003-0326-8945}}
\email{dawrasimran27@gmail.com}
\affiliation{Center for Gravitational Physics and Quantum Information, Yukawa Institute for Theoretical Physics, Kyoto University, 606-8502, Kyoto, Japan.
}%
\author{Sai Swagat Mishra\orcidlink{0000-0003-0580-0798}}
\email{saiswagat009@gmail.com}
\affiliation{SR University, Warangal, Telangana, 506371
}%
\affiliation{Department of Astronomy, University of Science and Technology of China, Hefei 230026, China}
\affiliation{School of Astronomy and Space Science, University of Science and Technology of China, Hefei 230026, China}%

\author{P. K. Sahoo\orcidlink{0000-0003-2130-8832}}
\email{pksahoo@hyderabad.bits-pilani.ac.in}
\affiliation{Department of Mathematics, Birla Institute of Technology and Science, Pilani, Hyderabad Campus, Jawahar Nagar, Kapra Mandal, Medchal District, Telangana 500078, India}

\begin{abstract}
\nolinenumbers
We investigate the late-time cosmological dynamics of a spatially flat FLRW Universe in the framework of exponential \(f(Q)\) gravity, where \(Q\) denotes the nonmetricity scalar. The matter sector is described by a bulk viscous fluid with a generalized barotropic equation of state characterized by a free parameter \(\gamma\), extending the standard pressureless matter scenario. We derive the modified background evolution equations and investigate the resulting cosmological dynamics. The model parameters are constrained using a combination of cosmic chronometer (CC), DESI DR2 baryon acoustic oscillation (BAO), CMB, and Type Ia supernova observations. Using the best-fit parameters, we analyze the phase-space dynamics of the model and employ the statefinder and \(\mathrm{Om}(z)\) diagnostics to quantify deviations from the standard \(\Lambda\)CDM cosmology. We further examine the evolution of the deceleration parameter and the effective equation-of-state parameter, demonstrating a transition from an early matter-dominated decelerating epoch to the present phase of accelerated expansion. Overall, the model provides a viable framework for late-time cosmic acceleration, recovering the standard matter-dominated evolution at early times while naturally evolving toward a stable late-time de Sitter attractor consistent with current observations.\\ 

\end{abstract}

\maketitle
\section{Introduction}
Observations over the past few decades have established a coherent picture of the late-time Universe as undergoing accelerated expansion. This conclusion is primarily supported by measurements from Type Ia supernovae (SNeIa)~\cite{riess1998observational,perlmutter1999constraining}, large-scale structure surveys~\cite{daniel2008large,koivisto2006dark}, baryon acoustic oscillations (BAO)~\cite{seo2003probing,blake2003probing,DESI:2024mwx,DESI:2025zgx} and cosmic microwave background (CMB)~\cite{Planck:2018vyg}.

Within the framework of general relativity (GR), the most standard explanation for this phenomenon is the presence of a cosmological constant, described by a constant and negative equation-of-state (EoS) parameter. This interpretation underlies the standard $\Lambda$CDM model, which has achieved notable success in reproducing a wide range of late-time observational data. Despite this success, the cosmological constant is plagued by a serious theoretical difficulty: its observed value is many orders of magnitude smaller than that predicted by quantum field theory and the requirement of finely tuned initial conditions \cite{Zlatev:1998tr,Weinberg:1988cp}. These shortcomings provide strong motivation to explore alternatives to a strictly constant dark-energy component. In this context, modifications and extensions of GR offer a natural framework in which cosmic acceleration can emerge dynamically at late times, while remaining compatible with constraints from low-redshift observations.

Amongst studies aiming to provide an alternative explanation for the effects of dark energy, modified theories based on curvature ($f(R)$ gravity) and torsion ($f(T)$ gravity and its extensions) have received special attention in recent years \cite{capozziello2008cosmography,sotiriou2010f,de2010f,harko2011f,shabani2014cosmological,cai2016f,linder2010einstein,Mishra:2025rhi}.  In light of the geometrical nature of gravity, it is essential to investigate the various geometrical forms that gravity might assume. The nonmetricity is the other fundamental characteristic connected to the connection, in addition to curvature and torsion. It has been suggested to create conditions where the gravitational interaction is mediated by nonmetricity, whereas curvature and torsion are vanishing, among other strong candidates for modified gravity theories. This gives us symmetric teleparallel gravity, further extended to $f(Q)$ gravity \cite{zhao2022covariant,PhysRevD.106.043509}. In contrast to $f(R)$ gravity fourth-order field equations, $f(Q)$ gravity has second-order field equations unaffected by pathologies. Implementing a flat connection in the presence of affine coordinates, where all its components vanish, transforming covariant derivatives into partial derivatives, is a crucial feature of the $f(Q)$ theory. Thus, the $f(Q)$ approach can distinguish gravity and inertial effects. As a result, the formulation of the $f(Q)$ theory serves as a new basis for several modified theories of gravity. 

Numerous investigations of modified \(f(Q)\) gravity have explored its cosmological and astrophysical implications, ranging from background cosmology to compact objects and late-time cosmic dynamics \cite{jimenez2020cosmology,d2022black,mandal2020cosmography,albuquerque2022designer,capozziello2022model,lymperis2022late,frusciante2021signatures,Gadbail:2024als,Kolhatkar:2024oyy,Sokoliuk:2023ccw,Arora:2022mlo}. In parallel, a variety of observational studies have constrained different \(f(Q)\) parameterizations using current cosmological datasets \cite{atayde2021can,lazkoz2019observational,d2022forecasting,ayuso2021observational,barros2020testing,anagnostopoulos2021first}. Despite these advances, the properties of the cosmic matter sector remain an important ingredient in determining the late-time expansion history, motivating investigations beyond the standard pressureless matter description.

In this context, dissipative processes provide a natural extension of the cosmic fluid. Bulk viscosity, in particular, generates an effective negative pressure capable of modifying the expansion dynamics and has therefore been widely studied as a mechanism for late-time cosmic acceleration. Within a relativistic hydrodynamic description, we adopt Eckart's formalism, which provides a phenomenological treatment of bulk viscous dissipation arising when the cosmic fluid departs from local thermodynamic equilibrium \cite{eckart1940thermodynamics}. Eckart's approach has been widely employed in modeling late-time cosmic acceleration driven by viscous effects \cite{ren2006modified}. Although Eckart's first-order theory is known to suffer from causality and stability issues, its mathematical simplicity has made it a widely adopted phenomenological framework for studying viscous cosmology. A causal formulation was subsequently developed by Israel and Stewart through a second-order relativistic theory incorporating a finite relaxation time for dissipative processes \cite{Israel:1979wp}. Despite its limitations, Eckart's formalism remains a useful approximation for investigations of the homogeneous background evolution and has been extensively employed in studies of bulk viscous cosmology. Within both approaches, bulk viscosity has been investigated as an effective mechanism for driving late-time cosmic acceleration and reproducing dark-energy-like behaviour in a variety of gravitational theories \cite{Avelino:2025lki,Palma:2025qge,singh2014friedmann,solanki2021cosmic,arora2022bulk,brevik2006crossing,Villalobos:2025mdk}.

In the present work, we adopt Eckart's first-order formalism to model bulk viscosity in the cosmic fluid. Within this framework, a bulk viscous fluid is characterized by the energy density $\rho$ and an effective pressure $p_{eff} = p + \Pi$ where $p$ denotes the equilibrium pressure and $\Pi$ represents the bulk viscous pressure. For a homogeneous and isotropic Universe, Eckart's theory yields $\Pi = -\zeta \,\vartheta$, $\zeta$ being the bulk viscosity coefficient and $\vartheta = 3H $ the expansion scalar, where $H=\frac{\dot{a}}{a}$ is the Hubble parameter. This formulation provides an effective description of dissipative effects at the background level and serves as the basis for our analysis in both general relativity and modified theories. We analyze the background evolution of a spatially flat Universe sourced by a bulk viscous fluid obeying the EoS $p=(\gamma-1)\rho$, within the frameworks of general relativity and $f(Q)$ gravity. We examine the impact of varying $\gamma$ on the cosmological dynamics and confront the resulting expansion histories with recent observational data, including observational Hubble measurements known as cosmic chronometers, type Ia supernovae, baryon acoustic oscillation data from DESI DR2, and compressed CMB. 

This paper is organized as follows. In Sec.~\ref{section 2}, we briefly review the theoretical framework of \(f(Q)\) gravity. In Sec.~\ref{section 3}, we derive the cosmological field equations and the Hubble expansion for both the viscous GR and exponential \(f(Q)\) models in the presence of a bulk viscous fluid. The observational methodology and dataset combinations are presented in Sec.~\ref{section 5}, while the corresponding observational constraints and reconstructed cosmological evolution are discussed in Sec.~\ref{section:result}. In Sec.~\ref{section 6}, we investigate the evolution of the deceleration parameter, the effective equation of state, and the statefinder and \(Om\) diagnostics. Finally, Sec.~\ref{section 7} summarizes our main results and conclusions.

\section{Overview of $f(Q)$ gravity}
\label{section 2}

To explore the cosmological implications of symmetric teleparallel gravity, we consider the most general form of the affine connection, following \cite{Hehl:1976kj,zhao2022covariant}
\begin{eqnarray}
\tilde{\Gamma}^{\alpha}_{\,\, \mu\nu}=  \mathring{\Gamma}^{\alpha}_{\,\,\mu\nu} + K^{\alpha}_{\,\,\mu \nu} +L^{\alpha}_{\mu \nu},
\end{eqnarray}
where the Levi-Civita connection of the metric $g_{\mu \nu}$, the contortion tensor, and disformation tensor are
\begin{eqnarray}
\mathring{\Gamma}^{\alpha}_{\,\,\mu\nu} &=& \frac{1}{2} g^{\alpha \sigma} \left(\partial_{\mu} g_{\sigma \nu} + \partial_{\nu} g_{\sigma \mu}- \partial_{\sigma} g_{\mu \nu}\right),\\
K^{\alpha}_{\,\,\mu \nu} &=&  \frac{1}{2} g^{\alpha \sigma} \left( - T_{\mu \sigma \nu} - T_{\nu \sigma \mu}+T_{\sigma \mu \nu}\right),\\
L^{\alpha}_{\,\,\mu \nu} &=& \frac{1}{2} g^{\alpha \sigma} \left( - Q_{\mu \sigma \nu}- Q_{\nu \sigma \mu} + Q_{\sigma \mu \nu}\right).
\end{eqnarray}
Here, the torsion and nonmetricity tensors are defined by 
\begin{eqnarray*}
T^{\alpha}_{\,\,\mu \nu} &= & \tilde{\Gamma}^{\alpha}_{\,\,\mu\nu} -\tilde{\Gamma}^{\alpha}_{\,\,\nu\mu},\\
Q_{\alpha \mu \nu} &= & \Delta_{\alpha} g_{\mu \nu} =  \partial_{\alpha} g_{\mu \nu} - \tilde{\Gamma}^{\lambda}_{\, \,  \mu \alpha} g_{\lambda \nu} - \tilde{\Gamma}^{\lambda}_{\,\, \nu \alpha} g_{ \mu \lambda}.
\end{eqnarray*}
Thus, the nonmetricity scalar is constructed as 
\begin{equation}
Q=-g^{\mu \nu} \left( L^{\alpha}_{\,\,\beta \mu} L^{\beta}_{\,\,\nu \alpha}- L^{\alpha}_{\,\,\beta \alpha} L^{\beta}_{\,\,\mu \nu} \right).
\end{equation}
Imposing the condition of symmetric teleparallelism renders the most general affine connection purely inertial. In this case, the connection can be written as
\begin{equation*}
  \tilde{\Gamma}^{\alpha}_{\,\, \mu\nu}= \frac{\partial x^{\alpha} }{\partial \xi^{\sigma}} \frac{\partial^2 \xi^{\sigma} }{\partial x^{\mu} \partial x^{\nu}}, 
\end{equation*}
where $\xi^{\sigma}$ denotes an arbitrary function of the spacetime coordinates. By performing an appropriate general coordinate transformation, one may always choose $x^{\alpha}=\xi^{\sigma}$, in which case the affine connection vanishes identically, $\tilde{\Gamma}^{\alpha}_{\,\, \mu\nu}=0$. This special choice of coordinates is known as the coincident gauge \cite{jimenez2018coincident,dialektopoulos2019noether,ayuso2021observational}. 
In the coincident gauge, the nonmetricity tensor reduces to $Q_{\alpha \mu \nu}= \partial_{\alpha}g_{\mu \nu}$, implying that covariant derivatives coincide with ordinary partial derivatives. As a consequence, within symmetric teleparallel geometry, the dynamics of the Einstein-Hilbert theory are exactly recovered from the nonmetricity-based Lagrangian $2\mathcal{L}=Q$.

We now extend symmetric teleparallel gravity by promoting the nonmetricity scalar $Q$ to an arbitrary function $f(Q)$. The corresponding action is given by
\begin{equation} 
\label{1}
S= \int \left[ -\frac{1}{2} f(Q)+L_{m} \right] \sqrt{-g} d^{4}x,
\end{equation}
where $L_{m}$ denotes the matter Lagrangian, $g$ is the determinant of the metric $g_{\mu \nu}$, and $f(Q)$ is a general function of the nonmetricity scalar. In addition, the two independent traces of the nonmetricity tensor and the nonmetricity scalar are defined as $Q_{\alpha} = g^{\mu \nu} Q_{\alpha \mu \nu}$,  $\tilde{Q}_{\alpha} =g^{\mu \nu} Q_{ \mu \alpha \nu}$ and $Q = -Q_{\alpha \mu \nu} P^{\alpha \mu \nu}$. 
The latter expression includes the nonmetricity conjugate 
\begin{equation}
\label{2}
P^{\alpha}_{\mu \nu}= -\frac{1}{2} L^{\alpha}_{\mu \nu} + \frac{1}{4} (Q^{\alpha}-\tilde{Q}^{\alpha})g_{\mu \nu}-\frac{1}{4} \delta^{\alpha}_{\,\,(\mu}\tilde{Q}_{\nu)}.
\end{equation}
Varying the action with respect to the metric and connection, respectively, we obtain the field equations of $f(Q)$ gravity
\begin{eqnarray}  
\label{4}
\nonumber
\frac{2}{\sqrt{-g}} \nabla_{\alpha}(\sqrt{-g} f_{Q} P^{\alpha}_{\,\,\,\,\mu \nu} )+\frac{1}{2} g_{\mu \nu} f + f_{Q} (P_{\mu \alpha \beta} Q_{\nu}^{\,\,\,\,\alpha \beta}\\
 - 2Q_{\alpha \beta \mu} P^{\alpha \beta}_{\,\,\,\,\,\,\, \nu})= T_{\mu \nu},
\end{eqnarray}
where $f_{Q}=\frac{df}{dQ}$. Here, $T_{\mu \nu}$ is the stress-energy momentum tensor  given by 
\begin{equation} 
\label{5}
T_{\mu \nu}= -\frac{2}{\sqrt{-g}} \frac{\delta(\sqrt{-g} L_{m})}{\delta g^{\mu \nu}}.
\end{equation}
The connection equation of motion can be derived by noting that the variation of the affine connection with respect to $\xi^{\alpha}$ is equivalent to an infinitesimal diffeomorphism. Consequently, one finds $\partial_{\xi} \tilde{\Gamma}^{\alpha}_{\,\, \mu\nu}=-\mathcal{L}_{\xi}  \tilde{\Gamma}^{\alpha}_{\,\, \mu\nu} = -\nabla_{\mu} \nabla_{\nu} \, \xi^{\alpha}$ \cite{jimenez2020cosmology,jimenez2018teleparallel}, where $\mathcal{L}_{\xi}$ denotes the Lie derivative. Furthermore, in the absence of hypermomentum, the field equations for the connection follow directly from varying the action \eqref{1} with respect to the connection
\begin{equation*}
    \nabla_{\mu}\nabla_{\nu}\left( \sqrt{-g} f_{Q} P^{\mu \nu}_{\,\,\, \,\,\, \alpha} \right)=0.
\end{equation*} 

\section{Cosmological formulation}
\label{section 3}
We start with the Friedmann-Lema\^{i}tre-Robertson-Walker (FLRW) metric in a homogeneous and isotropic flat spacetime
\begin{equation} 
\label{8}
ds^{2}= -dt^{2}+a^{2}(t)\delta_{ij}dx^{i}dx^{j},
\end{equation}
where $a(t)$ is the scale factor of the universe. The nonmetricity reduces to $Q=6 H^{2}$. 

Focusing our attention on the coincidence gauge, we can write
the modified Friedman equations as
\begin{eqnarray} 
\label{10}
&6 f_{Q} H^{2}-\frac{1}{2} f = \rho,\\
\label{11}
&(12 H^2 f_{QQ}+f_{Q})\dot{H}= -\frac{1}{2}(\rho+p),
\end{eqnarray}
where $f_{Q}=\frac{df(Q)}{dQ}$, $f_{QQ}=\frac{d^{2}f(Q)}{dQ^{2}}$. 
Here, \(\rho=\rho_m+\rho_z\) denotes the total energy density, consisting of pressureless matter \(\rho_m\) and a viscous fluid \(\rho_z\), while the total pressure is provided entirely by the viscous component.

The idea of a stiff fluid with EoS $p=\rho$, originally proposed by Zel'dovich, has long served as a useful theoretical benchmark in cosmology. This component dilutes rapidly ($\rho \propto a^{-6}$) and therefore does not affect late-time evolution in its standard non-viscous form. However, when bulk viscosity is included, the effective pressure can deviate significantly from the ideal stiff limit, modifying the cosmic dynamics in a nontrivial manner \cite{mathew2014cosmology,Zel/1962,Nair/2016}. This generalized viscous extension provides a broader phenomenological framework that interpolates between stiff, dust-like, and accelerated regimes, thereby motivating the study of viscous cosmologies within both GR and modified gravity scenarios.

We assume that the pressureless matter satisfies \(p_m=0\), while the viscous fluid obeys the generalized barotropic equation of state
\begin{equation}
p_z= (\gamma-1)\rho_z. 
\label{12}
\end{equation}
Here, $\gamma=1$ corresponds to pressureless dust and $\gamma=2$ reproduces the Zel'dovich stiff-fluid limit. The bulk viscosity modifies only the viscous-fluid pressure according to
\begin{equation}
\label{13}
p_z^{eff}= p_z-3 \zeta H, 
\end{equation}
where $\zeta$ denotes the bulk viscosity coefficient. Assuming that pressureless matter and the viscous fluid are separately conserved, their conservation equations are
\begin{equation}
\label{14}
\dot{\rho_m}+ 3H \rho_m=0, \quad \quad 
\dot{\rho_z}+ 3H\left(\rho_z +p_z^{eff}\right)=0.
\end{equation}
Consequently, the matter density evolves as \(\rho_m=\rho_{m0}a^{-3}\), while the evolution of the viscous component is governed by the effective pressure above.

\subsection{Viscous fluid dominated universe}

Before turning to the effects of the $f(Q)$ formulation of gravity with the EoS given in Eq.~\eqref{12}, it is instructive to first examine the dynamics GR. In this case, the modified-gravity contributions are absent, and the background evolution is governed by the standard Friedmann equation $3H^2=\rho$. The resulting analytical solution provides a useful reference against which the cosmological evolution in the subsequent $f(Q)$ model can be compared.

Using the conservation equations
\begin{eqnarray}
\dot{\rho}_m+3H\rho_m=0,\\
\dot{\rho}_z+3H(\rho_z+p_z)=0,
\end{eqnarray}
together with the effective EoS in Eq.~\eqref{12}, one obtains
\begin{equation}
\frac{dH}{dN}
=
-\frac{3}{2H}
\left[
(1-\gamma)\rho_m
+\gamma H^2
-\zeta H
\right],
\end{equation}
or, after introducing the dimensionless expansion rate $E\equiv H/H_0$ and using
\begin{equation}
\rho_m=3H_0^2\Omega_{m0}e^{-3N},
\end{equation}
the evolution equation becomes
\begin{equation}
\label{eq:GR_background}
\frac{dE}{dN}
=
-\frac{3}{2E}
\left[
(1-\gamma)\Omega_{m0}e^{-3N}
+\gamma E^2
-\frac{\zeta}{H_0}E
\right].
\end{equation}
This GR solution provides the reference background against which the modifications introduced by the exponential $f(Q)$ theory will be assessed.
In the limit $\tilde{\zeta} \equiv\frac{\zeta}{H_0}=0$, the viscous fluid reduces to a perfect fluid with barotropic equation of state $p=(\gamma-1)\rho$, recovering the standard dust ($\gamma=1$) and stiff-fluid ($\gamma=2$) limits. A non-vanishing bulk viscosity introduces an effective negative pressure, which can significantly modify the late-time expansion history. The dust case ($\gamma=1$) has been investigated extensively in the context of viscous cosmology, where it was shown that bulk viscous effects can mimic dark-energy behaviour and drive late-time cosmic acceleration \cite{Avelino:2008ph}. The Zel'dovich fluid ($\gamma=2$) with bulk viscosity has also been studied in Refs.~\cite{Mathew:2014gpa,Nair:2015bhz}, demonstrating that viscosity can transform an otherwise decelerating stiff-fluid universe into one exhibiting accelerated expansion. Motivated by these earlier studies, we consider the more general case in which the equation-of-state parameter is allowed to vary over the interval $0<\gamma\le2$.

\subsection{Viscous $f(Q)$ formalism}

In this section, we start our analysis using the exponential model, which tends to the $\Lambda$CDM paradigm at high redshifts while producing a non-negligible impact in low-$z$ cosmology. The $f(Q)$ function is given by
\begin{equation}
\label{17}
    f(Q)=Qe^{\alpha Q_0/Q},
\end{equation}
with $\alpha$ being a dimensionless constant, while $Q_0=6H_0^2$. Moreover, since the model tends to the GR limit at early times ($Q>Q_{0}$), it satisfies the BBN constraints. Note that the model effectively simplifies to the polynomial case at some point in cosmic history as $Q_{0}/Q$ declines. This model behaves like an endless variety of models coexisting, with one acting dominantly at each point in cosmic history. This functional form was originally proposed in the context of teleparallel $f(T)$ gravity \cite{Awad:2017yod} and was subsequently investigated within the framework of $f(Q)$ gravity \cite{Anagnostopoulos:2021ydo,Khyllep:2022spx,Anagnostopoulos:2022gej}. Since both the torsion scalar $T$ and the non-metricity scalar $Q$ are proportional to $H^2$ at the background level in a spatially flat FLRW universe, the corresponding cosmological background evolution of the exponential $f(T)$ and $f(Q)$ models is expected to be closely related, although the underlying geometric descriptions are fundamentally different.

The energy density of the viscous fluid, denoted by \(\rho_z\), can be obtained after decomposing the total energy density as \(\rho=\rho_m+\rho_z\), where \(\rho_m\) denotes the pressureless matter component. Evaluating Eq.~\eqref{10} at the present epoch together with the modified Friedmann equation gives the constraint
\begin{equation}
\label{19}
\alpha=\frac{1}{2}
+\mathcal{W}\left(
-\frac{\Omega_{m0}+\Omega_{z0}}
{2e^{1/2}}
\right),
\end{equation}
where \(\mathcal{W}(x)\) denotes the Lambert \(W\)-function, while \(\Omega_{m0}\) and \(\Omega_{z0}\) represent the present-day density parameters of pressureless matter and the viscous fluid, respectively. Consequently, the parameter \(\alpha\) is not independent but is completely determined once the present-day matter and viscous-fluid density parameters are specified. To determine the background evolution, we solve the modified Friedmann equations together with the conservation equations for pressureless matter and the viscous fluid. The pressureless matter satisfies $\dot{\rho}_m+3H\rho_m=0$. Using Eq.~\eqref{11} together with the effective pressure defined in Eq.~\eqref{13}, the evolution equation for the dimensionless Hubble parameter becomes
\begin{eqnarray}
\label{20}
E'=-\frac{3}{2}\left[\frac{(1-\gamma)\Omega_{m0}e^{-3N}
+\gamma\left(E^2-2\alpha\right)e^{\alpha/E^2}-\tilde{\zeta}E}
{E\left(1+\frac{\alpha}{E^2}\left(\frac{2\alpha}{E^2}-1\right)\right)e^{\alpha/E^2}}\right],
\end{eqnarray}
where $\tilde{\zeta}\equiv\frac{\zeta}{H_0}$. Due to the nonlinear structure introduced by the exponential $f(Q)$ function and the viscous contribution, this equation does not admit a closed-form analytical solution and is therefore solved numerically. Initial conditions are imposed at the present epoch, $E(z=0)=1$, and the resulting evolution is used to reconstruct the background expansion history.


\subsection{Phase portrait of exponential $f(Q)$ gravity}

We now investigate the late-time phase-space dynamics of the exponential \(f(Q)\) cosmological model. Since the background evolution Eq.~\eqref{20} contains the explicit matter contribution \(\Omega_{m0}e^{-3N}\), the full cosmological system is non-autonomous and is solved numerically in the previous subsection. However, at late times the pressureless matter density dilutes as \(a^{-3}\), so that \(\Omega_{m0}e^{-3N}\rightarrow0\). In this asymptotic regime the evolution equation becomes autonomous and depends only on the Hubble parameter. The resulting autonomous system provides a convenient framework for identifying the equilibrium points and studying the qualitative late-time dynamics of the model. Taking the limit \(\Omega_{m0}e^{-3N}\rightarrow0\) in Eq.~\eqref{20}, we obtain
\begin{eqnarray}
\label{doth}
\dot{H} = \frac{3}{2}
\frac{\left[\zeta H e^{-\alpha H_0^2/H^2}
-\gamma \left(H^2-2\alpha H_0^2\right)\right]H^4}
{H^4-\alpha H_0^2H^2+2\alpha H_0^4} = \mathcal{F}(H).
\end{eqnarray}
Since Eq. \eqref{doth} represents the late-time limit of the background evolution and depends only on \(H\), it defines a one-dimensional autonomous dynamical system that governs the asymptotic cosmological evolution. The expanding and contracting cosmological branches correspond to $H>0$ and $H<0$, respectively, while the sign of $\dot H$ determines the direction of evolution in phase space. The equilibrium points of the system are obtained by imposing $\mathcal{F}(H)=0$. Provided that the denominator remains finite, the critical points satisfy
\begin{eqnarray}
H^4\left[\zeta H e^{-\alpha H_0^2/H^2}
-\gamma\left(H^2-2\alpha H_0^2\right)\right]=0.
\end{eqnarray}
One immediate solution is the Minkowski spacetime $H=0$, while the remaining equilibrium points satisfy the transcendental equation
\begin{equation}
\zeta H e^{-\alpha H_0^2/H^2}=\gamma\left(H^2-2\alpha H_0^2\right),
\label{fixedpoint}
\end{equation}
whose solutions correspond to de Sitter universes with constant Hubble expansion. Since Eq.~\eqref{fixedpoint} cannot be solved analytically, the corresponding fixed points are determined numerically for the observationally constrained parameter values. To gain further insight into the cosmological evolution, it is instructive to examine the asymptotic behaviour of the autonomous system. In the early-Universe limit ($H\gg H_0$), Eq.~\eqref{doth} becomes
\begin{equation}
\dot H = \frac32 \frac{\zeta H-\gamma H^2+\mathcal{O}(H_0^2)} {1-\alpha H_0^2/H^2+\mathcal{O}(H^{-4})}.
\end{equation}
To leading order in $H_0^2/H^2$ in the presence of viscosity, this reduces to $\dot H \simeq \frac32\left(\zeta H-\gamma H^2\right)$. Furthermore, for sufficiently large expansion rates ($H\gg\zeta/\gamma$), the quadratic term dominates over the viscous contribution, yielding $\dot H \simeq -\frac32\gamma H^2$. Using the barotropic relation $\gamma=1+w$, we recover the standard FLRW evolution, $\dot H = -\frac32(1+w)H^2$, showing that the exponential $f(Q)$ model naturally reproduces the conventional matter- and radiation-dominated epochs at early times.

In the opposite limit ($H\ll H_0$), the exponential contribution becomes exponentially suppressed $e^{-\alpha H_0^2/H^2}\rightarrow0$ and the autonomous equation reduces to
\begin{equation}
\dot H
\approx
-\frac32\gamma
\frac{(H^2-2\alpha H_0^2)H^4}
{H^4-\alpha H_0^2H^2+2\alpha H_0^4}.
\end{equation}
This asymptotic expression governs the late-time dynamics of the model. The corresponding late-time cosmological evolution is determined by the fixed points of Eq.~\eqref{doth}, whose number and stability depend on the model parameters $\alpha$, $\zeta$, and $\gamma$. These fixed points, together with their stability properties, will be analysed numerically in the following subsection through the corresponding phase portraits.

\begin{figure*}
\centering
\includegraphics[scale = 0.5]{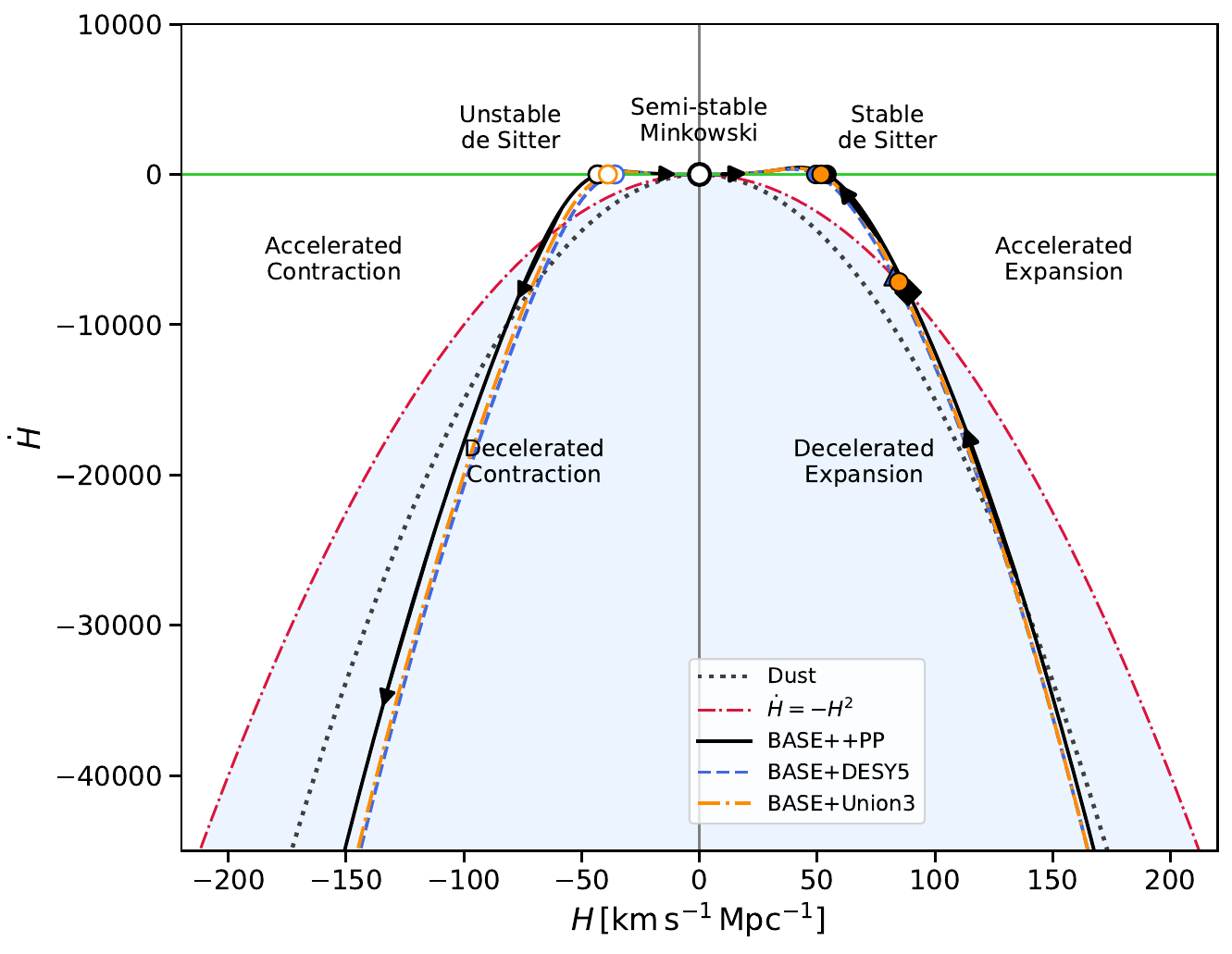}
\caption{Phase-space portraits of the exponential $f(Q)$ model reconstructed using the best-fit parameters from the three observational datasets: BASE+PP, BASE+DESY5, and BASE+Union3, where BASE denotes the combination CC+DESI+CMB. The dotted black curve represents the standard dust solution, $\dot H=-\frac{3}{2}H^2$, while the red dash-dotted curve denotes the zero-acceleration boundary, $\dot H=-H^2$.}
 \label{fig:phase}
 \end{figure*}

Fig.~\ref{fig:phase} presents the phase-space portrait of the exponential \(f(Q)\) model reconstructed using the best-fit parameters from the BASE+PP, BASE+DESY5, and BASE+Union3 observational datasets, where BASE denotes the combination of CC+DESI+CMB. The corresponding best-fit values of the equation-of-state parameter are \(\gamma=1.27\), \(\gamma=1.33\), and \(\gamma=1.33\), respectively. The horizontal axis denotes the Hubble parameter \(H\), while the vertical axis represents its time derivative \(\dot{H}\). For comparison, the dotted black curve shows the standard dust-dominated evolution, \(\dot{H}=-\frac{3}{2}H^2\), whereas the red dash-dotted curve represents the zero-acceleration boundary, \(\dot{H}=-H^2\), corresponding to \(q=0\). Consequently, the region above this curve describes an accelerating Universe (\(q<0\)), while the region below it corresponds to decelerated expansion (\(q>0\)).

The autonomous system admits two hyperbolic de Sitter fixed points together with a non-hyperbolic Minkowski solution at \(H=0\). The Minkowski point behaves as a semi-stable equilibrium, whereas the de Sitter solutions form an unstable repeller on the contracting branch (\(H<0\)) and a stable attractor on the expanding branch (\(H>0\)). Their stability is determined by the eigenvalue of the linearized system, with \(F'(H_*)>0\) indicating an unstable fixed point and \(F'(H_*)<0\) a stable attractor. For the BASE+PP dataset, the de Sitter solutions are located at \(H\simeq-43.29\) and \(H\simeq54.09~{\rm km\,s^{-1}\,Mpc^{-1}}\). The corresponding values are \(H\simeq-35.93\) and \(H\simeq49.43~{\rm km\,s^{-1}\,Mpc^{-1}}\) for BASE+DESY5, and \(H\simeq-38.89\) and \(H\simeq51.64~{\rm km\,s^{-1}\,Mpc^{-1}}\) for BASE+Union3. The negative de Sitter solution therefore acts as the past asymptotic state of contracting trajectories, whereas the positive de Sitter solution governs the late-time evolution of the expanding Universe.

The onset of cosmic acceleration is determined by the condition \(\dot{H}=-H_{\rm tr}^{\,2}\). The transition occurs at \(H_{\rm tr}\simeq88.60~{\rm km\,s^{-1}\,Mpc^{-1}}\) for BASE+PP, \(82.33~{\rm km\,s^{-1}\,Mpc^{-1}}\) for BASE+DESY5, and \(84.64~{\rm km\,s^{-1}\,Mpc^{-1}}\) for BASE+Union3. In every case, the transition precedes the approach to the stable de Sitter attractor, implying that the Universe evolves from a decelerating matter-dominated phase into an accelerating epoch before asymptotically reaching the de Sitter state. Although the transition scale and asymptotic expansion rate vary slightly among the three datasets, the overall dynamical evolution remains unchanged.

An important feature of Fig.~\ref{fig:phase} is that all three trajectories closely follow the dust solution at large values of \(H\), demonstrating that the exponential modification becomes negligible in the high-nonmetricity regime and naturally recovers the standard matter-dominated evolution. As the Hubble parameter decreases, the trajectories depart smoothly from the dust solution, cross the acceleration boundary, and converge to their respective stable de Sitter attractors. The persistence of the same phase-space topology across all three observationally favoured parameter sets indicates that the qualitative cosmological behaviour of the exponential \(f(Q)\) model is robust against the choice of late-time dataset.

\section{Cosmological constraints}
\label{section 5}

The next step of our analysis is to constrain the proposed models using observational data. To this end, we perform a MCMC analysis employing a combination of independent cosmological probes, including cosmic chronometers (CC), baryon acoustic oscillation measurements from DESI DR2, compressed CMB, and Type Ia supernova compilations from PantheonPlus, DES-Dovekie and Union3.

The parameter estimation is carried out using the publicly available Python package \texttt{emcee} \cite{foremanmackey2013emcee}, which implements an affine-invariant ensemble sampler well suited for exploring multidimensional parameter spaces. The resulting posterior distributions are subsequently analyzed and visualized using the \texttt{GetDist} package \cite{lewis2019getdist}, allowing for the computation of marginalized constraints, confidence intervals, and joint contour plots. In the following, we provide a detailed description of the observational datasets and outline the methodology adopted for likelihood construction and parameter inference.
\begin{itemize}
    \item Cosmic Chronometers (CC): The CC dataset is one of the most developed observational datasets that tests various cosmological models and is derived using the so-called differential age of the galaxies described in \cite{Yu_2018,2015MNRAS.450L..16M}. With the method mentioned above, one could derive the Hubble parameter value at a particular redshift with the following formula
\begin{equation*}
  H(z)=\frac{-1}{1+z}\frac{dz}{dt}.
\end{equation*}
One could derive the ratio using the equation $dz/dt$ as $\Delta z/\Delta t$, where $\Delta z$ can be thought of as the separation of a redshift in a dataset, values of which could be easily derived from the high-resolution spectroscopy. On the other hand, the derivation of $\Delta t$ is rather complex and challenging. $\Delta t$ usually could be derived from old stellar populations that endanger most galaxies up to very high redshifts.

The 31 data points considered in this study are collected from the Refs.~\cite{Jimenez:2003iv, Moresco:2012jh, Moresco:2015cya, Moresco:2016mzx}.
The methodology for constructing the $\chi^2$ function is discussed in detail in Ref.~\cite{Kavya:2025vsj}, and the full covariance matrix corresponding to the correlated CC dataset is taken from Ref.~\cite{Moresco:2020fbm}.
\item DESI BAO DR2: This data set consists of Baryon Acoustic Oscillation (BAO) measurements from the Dark Energy Spectroscopic Instrument (DESI) Data Release II~\cite{DESI:2025zgx}, which extends and improves upon the earlier DR1 results~\cite{DESI:2024mwx,Moon:2023jgl}. The key observables are the ratios \( \{D_M/r_d, D_H/r_d, D_V/r_d\} \), where \( D_M \) is the comoving angular diameter distance, \( D_H \) the Hubble distance, \( D_V \) the spherically averaged BAO distance, and \( r_d \) the comoving sound horizon at the drag epoch~\cite{eBOSS:2020yzd,DESI:2024mwx}. In recent years, such datasets have revealed intriguing hints that dark energy may evolve over time, challenging the standard $\Lambda$CDM model \cite{Wang:2025bkk, DESI:2025wyn, Mishra:2025goj, Arora:2025msq, Mishra:2026mnras}. A detailed description of DESI DR2 and the formulation of $\chi^2$ is presented in the Ref.~\cite{Mishra:2025vpy}.

\item SNeIa: Type Ia supernovae (SNeIa) are widely used as standard candles due to their relatively uniform intrinsic luminosity. This data set provides measurements of the apparent magnitude \( m_{b}(z) \), from which the luminosity distance \( D_{L}(z) \) is inferred via the magnitude-redshift relation
\begin{equation}
			\mu \equiv m-M = 5\log(D_L/\text{Mpc}) + 25 \  ,
		\end{equation}
		where, \(m\) denotes the apparent magnitude of the supernova and \(D_L\) is the luminosity distance: 	
		\begin{equation}
			D_L({z}) = c(1+{z}) \int_0^{{z}} \frac{dz'}{H(z')} \ ,
		\end{equation} 
assuming a flat FLRW metric, and \(c\) is the speed of light in km/s. The model parameters are constrained by minimizing the chi-square ($\chi^2$) likelihood, defined as:
		\begin{equation}
			-2 \ln (\mathcal{L}) = \chi^2 = \Delta D^{T} \mathcal{C}^{-1} \Delta D_j\ ,
		\end{equation}
		where $\Delta D = \mu_{ Obs} - \mu_{Model}$, $C^{-1}$ denotes the inverse combined statistical and systematic covariance matrix of the SNe sample. We use three different SNe datasets, including Pantheon+ \cite{Brout_2022}, DESY5 \cite{DES:2024jxu,DES:2024hip} and Union3 \cite{Rubin:2023jdq}.
        
\textbf{PP Data:} This data set refers to the Pantheon+ compilation, which includes 1550 spectroscopically confirmed Type Ia supernovae. The catalog provides 1770 data samples, from which we use the observational column corresponding to the non-SH0ES-calibrated apparent magnitude \( m_{\rm obs} \). We denote this subset as ``PP'' throughout our analysis.
		
\textbf{DESY5 Data:} This data set consists of Type Ia supernovae observations from the Dark Energy Survey five-year sample (DES-SN5YR), comprising 1829 distinct SNe. It includes 194 nearby SNe with redshift \( z < 0.1 \) and 1635 DES SNe. For our analysis, we compute the likelihood using the distance modulus \( \mu \) and the full covariance matrix provided in the data release. We have updated our supernova analysis to incorporate the DES-Dovekie release, which replaces the earlier DESY5 sample. The DES-Dovekie compilation includes 197 low-redshift SNe~Ia and 1623 DES likely SNe~Ia, with improved calibration and systematic control, making it approximately more robust than the previous dataset.

\textbf{Union3 Data:} The Union3 compilation, produced by the Supernova Cosmology Project \cite{Rubin:2023jdq}, provides one of the most extensive and homogeneous samples of Type~Ia supernovae to date. It extends the earlier Union and Union2 releases with improved photometric calibration, refined systematic corrections, and a uniformly processed sample of more than 1400 spectroscopically confirmed SNe Ia spanning the redshift range \( z \simeq 0.01\text{--}2.26 \). The dataset offers critical constraints on the cosmic expansion history and the properties of dark energy. Union3 combines observations from multiple surveys, including SNLS, SDSS, Pan-STARRS, CSP, and several low-redshift programs, along with space-based data from the Hubble Space Telescope (HST). All light curves were reprocessed using a common SALT2 pipeline to ensure photometric consistency.

\item CMB: This data set is derived by compressing the full Planck data set \cite{Planck:2018vyg} as reported in \cite{Chen:2018dbv}. The observables correspond to distance priors -- the acoustic scale \(l_A\) and shift parameter \(R\). These priors effectively encode information about the CMB temperature power spectrum. The acoustic scale characterizes the angular scale of the sound horizon, influencing peak spacing, while the shift parameter affects the line-of-sight direction, impacting peak heights. The shift parameter is defined as \cite{Elgaroy:2007bv}: 
	\begin{equation}
		R(z_*) \equiv \frac{D_A(z_*) \sqrt{\Omega_{m0} H_0^2}}{c}\ ,
	\end{equation}
	where \(z_* = 1089.92\) is the redshift at photon decoupling \cite{Planck:2018vyg}, and \(D_A\) is the angular diameter distance for a flat geometry given by:
	\begin{equation}
		D_A = c \int_{0}^{z} \frac{dz}{H(z)} \ .
	\end{equation}
	Here, \(\Omega_{m0}\) denotes the total matter density at present. For the minimally coupled case, where both baryonic and dark matter are pressureless and individually conserved, baryon density does not explicitly appear in the Hubble parameter \(H = H_0 \sqrt{\Omega_\Lambda + \Omega_{m0} (1+z)^3 + \Omega_r (1+z)^4}\). However, the total matter density is \(\Omega_{m0} = \Omega_{dm0} + \Omega_{b0}\), since the non-gravitational coupling only modifies the dark matter profile while baryons remain pressureless with \(\Omega_b \propto (1+z)^3\). Therefore, care must be taken when applying such models. The acoustic scale is defined as:
	\begin{equation}
		l_A = \frac{\pi D_A(z_*)}{r_s(z_*)}
	\end{equation}
	where \(r_s\) is the comoving sound horizon at the photon decoupling epoch. Along with these, the compressed CMB data includes \(\Omega_{b_0} h^2\). The data set includes the correlation matrix between \(\{R, l_A, \Omega_{b_0} h^2\}\), whose inverse covariance matrix is reported in the appendix of \cite{Chen:2018dbv}. The chi-square is calculated as:
	\begin{equation}
		\chi^2_{\rm PLA} = \sum \rm (D_i^{ obs}- D_i^{th}) C_{ij}^{-1} (D_j^{ obs}- D_j^{th}) \ .
	\end{equation}
 In the current study, we use the compressed dataset, which has been further improved in \cite{Arendse:2019hev,Cai:2014ela,Liu:2024fjy} by including an additional observable. The sample is extended to a four-parameter likelihood, $\{100,\Omega_b h^2, 100,\theta_{*}, R, \Omega_{\rm dm} h^2\}$, along with its correlation matrix. This extension has been shown to be efficient in constraining models beyond $\Lambda$CDM. The current likelihood replaces the acoustic scale with the angular scale defined as $\theta_{*} = r_s(z_*)/D_A(z_*)$ and introduces an additional likelihood parameter corresponding to the dark matter density $\Omega_{\rm dm}$. Additionally, we compute the redshift at recombination and at the photon-baryon decoupling epoch using the Hu-Sugiyama fitting formula \cite{Hu:1995en}.

\end{itemize}

\begin{figure*}
\centering
\includegraphics[width=1.0\columnwidth]{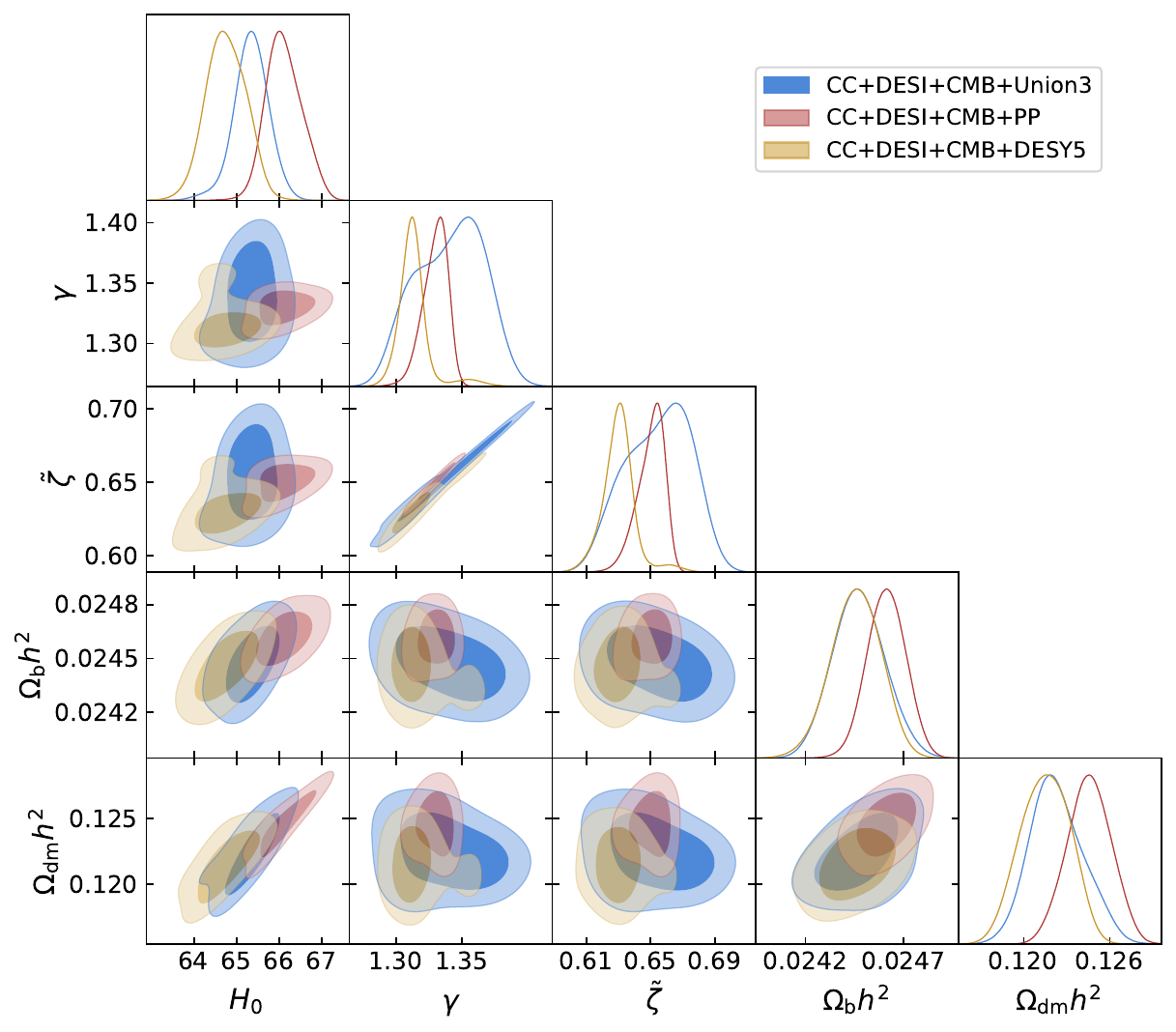}
\includegraphics[width=1.0\columnwidth]{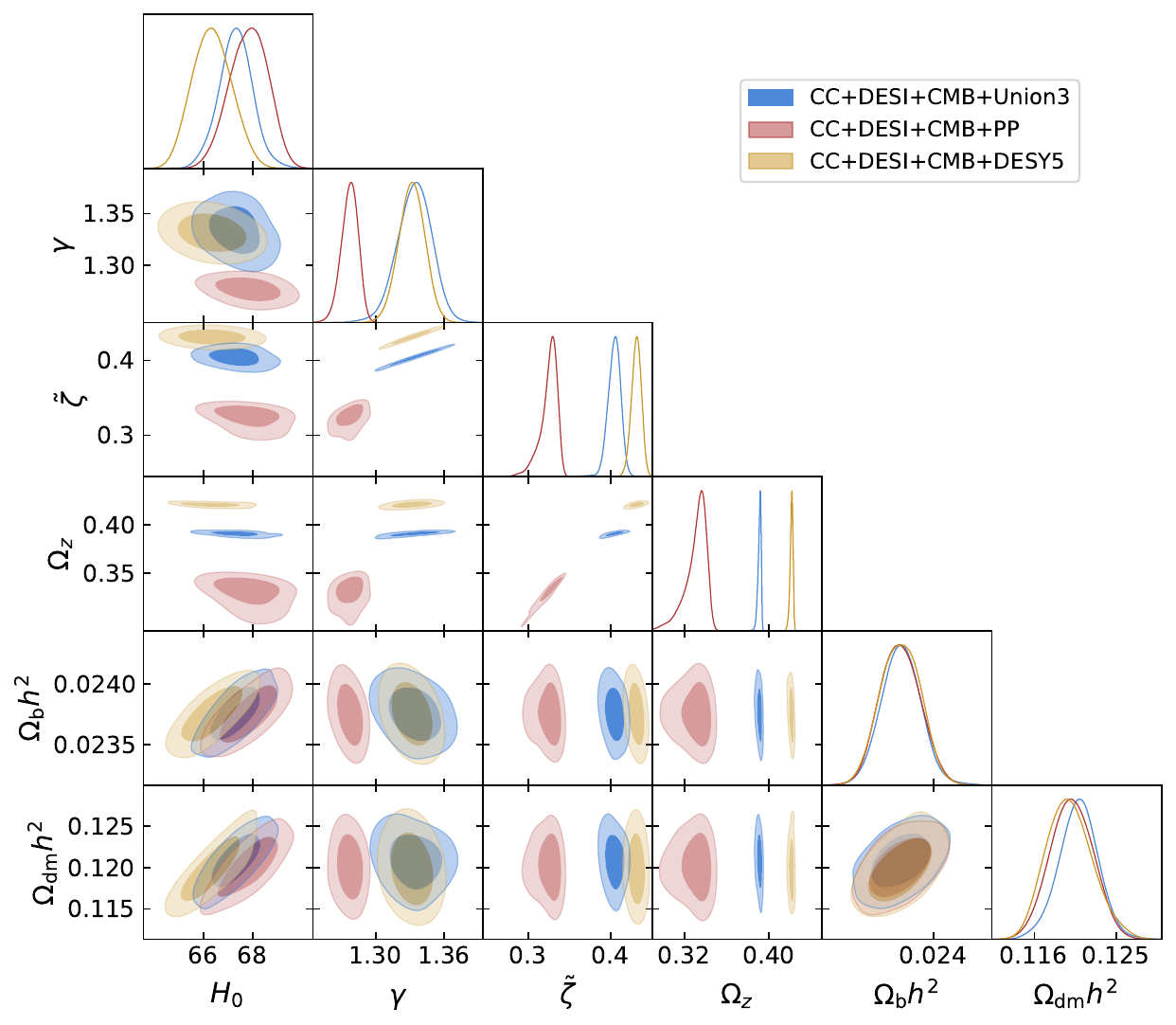}
\caption{\centering The $1\sigma$ and $2\sigma$ marginalized posterior distributions for the viscous GR model (left panel) and the viscous exponential $f(Q)$ model (right panel)}, obtained from the considered observational dataset combinations.
 \label{fig:contour}
 \end{figure*}

\section{Results} \label{section:result}

Having established the theoretical framework and numerical methodology, we now constrain the model parameters using cosmological observations. We perform a Bayesian parameter estimation for the exponential \(f(Q)\) model using the combinations BASE+PP, BASE+DESY5, and BASE+Union3.

The resulting marginalized constraints and best-fit parameter values are summarized in Table~\ref{tab:bestfit_models}. Throughout the analysis, the equation-of-state parameter \(\gamma\) is treated as a free parameter together with the bulk viscosity coefficient, allowing us to investigate the generalized viscous cosmology within both the GR and exponential \(f(Q)\) frameworks. Fig.~\ref{fig:contour} present the corresponding one- and two-dimensional posterior distributions for the different dataset combinations.

In the viscous GR model, the posterior distributions exhibit a clear positive correlation between the Hubble constant \(H_0\) and the bulk viscosity parameter \(\tilde{\zeta}\), indicating that larger viscous effects favour higher expansion rates. The equation-of-state parameter \(\gamma\) is mildly anti-correlated with both \(H_0\) and \(\tilde{\zeta}\), whereas the baryon and cold dark matter densities remain tightly constrained and only weakly correlated with the remaining parameters. The constraints obtained from the BASE+PP, BASE+DESY5, and BASE+Union3 analyses are mutually consistent, with only small shifts in the preferred parameter values.

The exponential \(f(Q)\) model exhibits a distinct posterior structure due to the additional geometric degree of freedom. In this case, \(\gamma\), \(\tilde{\zeta}\), and \(\Omega_{z0}\) are positively correlated, while \(\Omega_bh^2\) and \(\Omega_{\rm dm}h^2\) show mild negative correlations with the viscous-sector parameters. As in the viscous GR model, the matter densities are tightly constrained and remain strongly correlated with each other. The substantial overlap of the posterior contours for the three dataset combinations indicates that the inferred cosmological parameters are robust against the choice of supernova compilation.

 \begin{table*}
\centering
\caption{Best-fit values and constraints for the viscous-dominated and viscous exponential $f(Q)$ models obtained from different combinations of observational datasets.}
\label{tab:bestfit_models}
\vskip -3mm
\renewcommand{\arraystretch}{1.25}
\begin{tabular}{lcccccccc}
\hline\hline
\textbf{Model}
& $H_0\;[\mathrm{km\,s^{-1}\,Mpc^{-1}}]$
& $\tilde{\zeta}$
& $\gamma$ 
& $\Omega_z$
& $\Omega_b h^2$
& $\Omega_{c} h^2$\\
\hline
\multicolumn{7}{c}{\textbf{CC + DESI + CMB + PP}} \\
\hline
Viscous
& $66.06^{+0.49}_{-0.36}$
& $0.65^{+0.006}_{-0.010}$
& $1.33^{+0.0074}_{-0.010}$
& --
& $0.0246 \pm {0.00010}$
& $0.1245 \pm {0.0015}$
\\
$f(Q)$
& $67.89^{+0.74}_{-0.84}$
& $0.327^{+0.0074}_{-0.012}$
& $1.27^{+0.0064}_{-0.0079}$
& $0.334^{+0.0064}_{-0.0128}$
& $0.0237 \pm {+0.00015}$
& $0.12007^{+0.00238}_{-0.00238}$
 \\
\hline
\multicolumn{7}{c}{\textbf{CC + DESI + CMB + DESY5}} \\
\hline
Viscous
& $64.73^{+0.53}_{-0.43}$
& $0.630^{+0.006}_{-0.008}$
& $1.31^{+0.006}_{-0.007}$
& --
& $0.0244\pm{+0.00012}$
& $0.1215 \pm {0.00181}$
 \\
$f(Q)$ 
& $66.33^{+0.81}_{-0.80}$
& $0.431^{+0.0048}_{-0.0062}$
& $1.33 \pm {+0.0108}$
& $0.42^{+0.00116}_{-0.0014}$
& $0.0237 \pm {+0.00015}$
& $0.1197^{+0.0026}_{-0.0025}$
 \\
\hline
\multicolumn{7}{c}{\textbf{CC + DESI + CMB + Union3}} \\
\hline
Viscous
& $65.34 \pm 0.38$
& $0.65^{+0.018}_{0.026}$
& $1.344^{+0.023}_{-0.035}$
& --
& $0.0244^{+0.00013}_{-0.00012}$
& $0.1221^{+0.0021}_{-0.0015}$
\\
$f(Q)$
& $67.32^{+0.63}_{-0.65}$
& $0.404^{+0.0069}_{-0.0080}$
& $1.33^{+0.014}_{-0.015}$
& $0.391^{+0.0011}_{-0.0017}$
& $0.0237 \pm {+0.00014}$
& $0.12087^{+0.0020}_{-0.0021}$
 \\
\hline\hline
\end{tabular}
\end{table*}

\begin{figure*}
 \centering
  \includegraphics[width=0.49\linewidth]{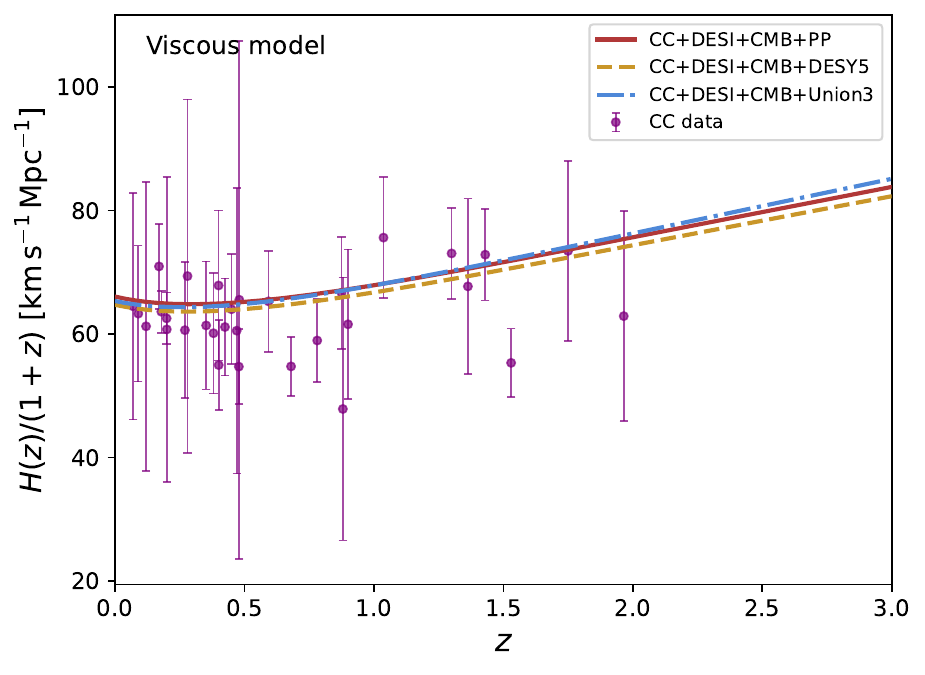}
  \includegraphics[width=0.49\linewidth]{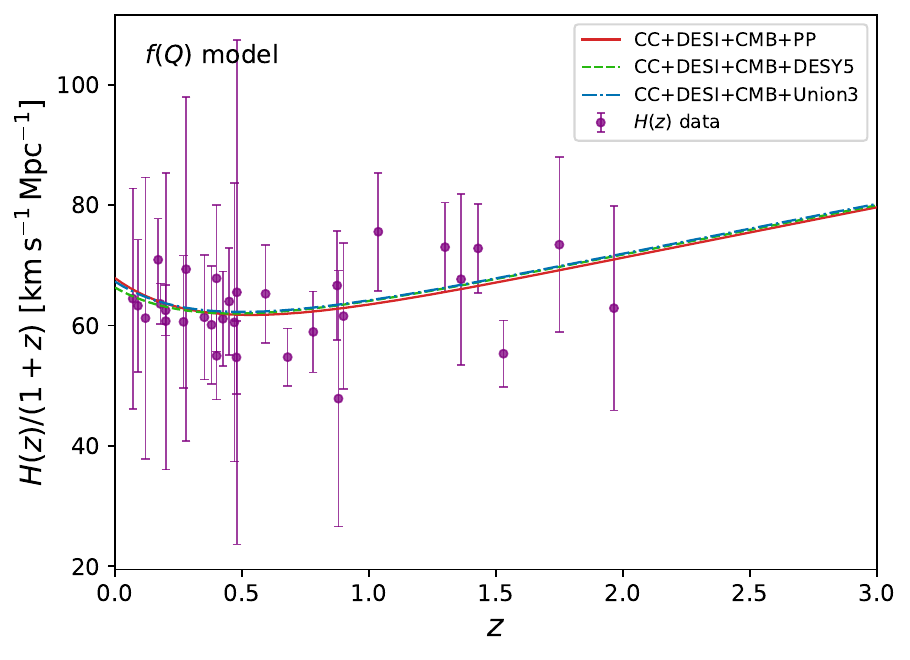}
 \caption{Best-fit reconstructed expansion histories, expressed as $H(z)/(1+z)$, for the viscous GR (left) and exponential $f(Q)$ (right) models. The curves correspond to the best-fit reconstructions obtained from the BASE+PP, BASE+DESY5, and BASE+Union3 dataset combinations.}
 \label{fig:hz}
 \end{figure*}

Fig.~\ref{fig:hz} shows the reconstructed expansion histories for the viscous GR model (left panel) and the exponential \(f(Q)\) model (right panel), expressed in terms of \(H(z)/(1+z)\). The reconstructions are obtained using the best-fit parameters from the BASE+PP, BASE+DESY5, and BASE+Union3 dataset combinations and are compared with the cosmic chronometer measurements. For both models, the three best-fit reconstructions closely overlap over the entire redshift range, demonstrating that the inferred expansion history is largely insensitive to the choice of supernova compilation. In each case, \(H(z)/(1+z)\) exhibits a shallow minimum around \(z\simeq0.4\), corresponding to the transition between the late-time accelerating phase and the earlier decelerating expansion. A comparison of the two panels shows that both the viscous GR and exponential \(f(Q)\) models provide nearly identical background expansion histories over the observed redshift range. The differences between the dataset combinations are small and become only marginally visible at intermediate and high redshifts.

To compare the performance of the viscous GR and viscous exponential $f(Q)$ models with the reference $\Lambda$CDM cosmology, we consider the minimum $\chi^2$ together with the Akaike Information Criterion (AIC) and the Bayesian Information Criterion (BIC), as summarized in Table~\ref{tab:model_comparison}. These are defined as
\[
\mathrm{AIC}=\chi^2_{\min}+2k, \qquad
\mathrm{BIC}=\chi^2_{\min}+k\ln N,
\]
where $k$ is the number of free parameters in the model and $N$ is the total number of independent observational data points entering the likelihood analysis. Model comparison is performed using the differences $\Delta\mathrm{IC}=\mathrm{IC}_{\rm model}-\mathrm{IC}_{\Lambda{\rm CDM}}$ computed relative to the corresponding $\Lambda$CDM fit~\cite{Akaike:1974vps}. $\Delta\mathrm{AIC}<2$ indicates substantial support for a model, $2\leq\Delta\mathrm{AIC}<6$ indicates moderate support, $6\leq\Delta\mathrm{AIC}<10$ indicates weak support, and $\Delta\mathrm{AIC}\geq10$ suggests that the model is strongly disfavoured \cite{Burnham2004,Liddle:2007fy,schwarz1978estimating}. For the BIC~\cite{Kass:1995loi, Trotta:2008qt, Arevalo:2016epc}, $0<\Delta\mathrm{BIC}<2$ corresponds to {\color{blue}evidence that is not worth more than a bare mention} against a model, $2\leq\Delta\mathrm{BIC}<6$ to positive evidence, $6\leq\Delta\mathrm{BIC}<10$ to strong evidence, and $\Delta\mathrm{BIC}\geq10$ to very strong evidence against the model relative to $\Lambda$CDM.

Table~\ref{tab:model_comparison} summarizes the statistical comparison of the viscous GR and viscous exponential $f(Q)$ models relative to the reference $\Lambda$CDM cosmology. For the BASE+PP dataset, the viscous exponential $f(Q)$ model yields the lowest $\chi^2_{\min}$ among the extended models and is marginally preferred over $\Lambda$CDM according to the AIC ($\Delta\mathrm{AIC}=-1.48$). However, the additional model complexity results in a substantially larger BIC penalty, with $\Delta\mathrm{BIC}=14.72$, indicating very strong evidence in favour of the simpler $\Lambda$CDM model. For the BASE+Union3 dataset, the viscous exponential $f(Q)$ model again provides a better fit than the viscous GR model. In this case, the AIC indicates moderate evidence against the extended models, whereas the BIC yields $\Delta\mathrm{BIC}=1.63$ for the exponential $f(Q)$ model, suggesting that it remains statistically comparable to the reference cosmology. For the BASE+DESY5 dataset, both the AIC and BIC favour the $\Lambda$CDM model, with the viscous GR model providing a slightly better fit than the exponential $f(Q)$ model among the two extended scenarios. Overall, although the information criteria generally favour the simpler $\Lambda$CDM model, both the viscous GR and exponential $f(Q)$ models remain compatible with the current observational data and provide competitive descriptions of the late-time cosmological expansion.\\

Finally, we examine the predictions of the viscous and $f(Q)$ models for the growth of matter perturbations. The linear growth rate is defined as
\begin{equation}
    f \equiv \frac{d\ln\delta}{d\ln a},
\end{equation}
where $\delta$ denotes the matter density contrast. In the $f(Q)$ scenario, the evolution of the growth rate can be expressed as~\cite{Khyllep:2022spx,Khyllep:2021pcu}
\begin{equation}
    \frac{df}{d\ln a}
    + f^2
    + \left(2+\frac{d\ln H}{d\ln a}\right)f
    =
    \frac{3}{2}\frac{\Omega_m(a)}{f_Q},
\end{equation}
where
\begin{equation}
    \Omega_m(a)
    =
    \frac{\Omega_{m0}a^{-3}}{E^2(a)},
    \qquad
    E(a)=\frac{H(a)}{H_0}.
\end{equation}
Here, $f_Q\equiv \partial f(Q)/\partial Q$ characterizes the modification of the gravitational interaction. 
Therefore, compared with the standard GR case, the growth of matter perturbations in the $f(Q)$ model is affected both by the modified background expansion and by the modification of the effective gravitational source term. The standard GR growth equation is recovered in the limit $f_Q\rightarrow 1$. For the viscous model, gravity remains governed by GR, and hence the corresponding source term is recovered by setting $f_Q=1$, while the viscous dynamics modify the growth history through the background expansion.

\begin{figure*}
 \centering
  \includegraphics[scale=0.5]{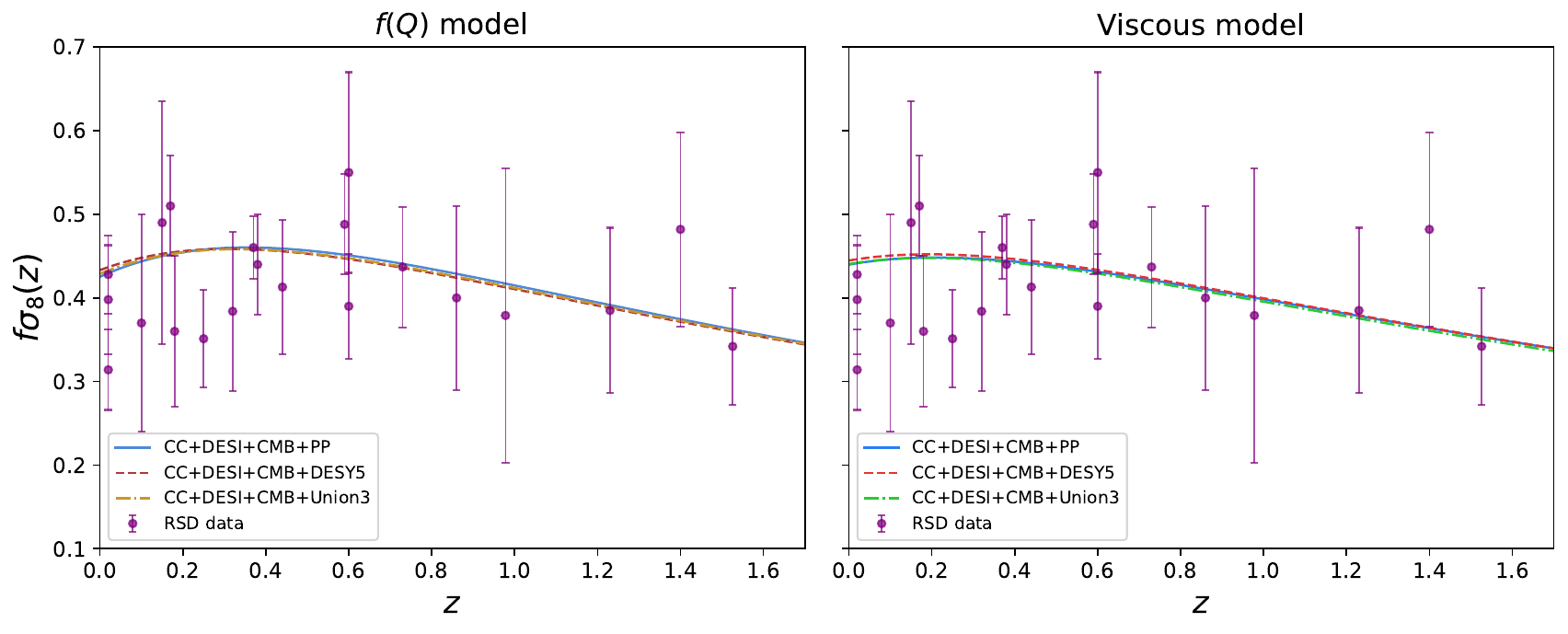}
 \caption{Evolution of the growth observable $f\sigma_8(z)$ for the $f(Q)$ and viscous models. The curves correspond to the best-fit model parameters obtained from the BASE+PP, BASE+DESY5, and BASE+Union3 dataset combinations. The observational measurements of $f\sigma_8(z)$ from redshift-space distortion data are shown with their corresponding $1\sigma$ uncertainties. A fixed value of $\sigma_{8,0}=0.81$ is adopted for all curves.}
 \label{fig:fsigma8}
 \end{figure*}

To confront these predictions with observations, we consider the quantity $f\sigma_8(z)$, which combines the linear growth rate with the root-mean-square amplitude of matter fluctuations, $\sigma_8(z)$, and can be directly constrained by redshift-space distortion (RSD) measurements~\cite{Alestas:2022gcg,Macaulay:2013swa,Nesseris:2017vor}. The resulting evolution of $f\sigma_8(z)$ for both models is presented in Fig.~\ref{fig:fsigma8}. We adopt a fixed present-day normalization of $\sigma_{8,0}=0.81$ for all dataset combinations, such that the differences among the predicted curves primarily arise from the distinct background and gravitational dynamics. The predictions are shown for the BASE+PP, BASE+DESY5, and BASE+Union3 dataset combinations.

As shown in Fig.~\ref{fig:fsigma8}, the predicted growth histories are broadly compatible with the available RSD measurements within their observational uncertainties. At the same time, differences are visible among the dataset combinations and between the viscous and $f(Q)$ scenarios, reflecting the sensitivity of structure growth to both the expansion history and the underlying gravitational dynamics. More precise measurements of the growth rate could therefore provide an important complementary probe for distinguishing these scenarios beyond constraints derived solely from the background expansion.

\section{Cosmological parameters}
\label{section 6}

To assess the physical viability of the cosmological models constructed in this work, it is useful to examine a set of diagnostic parameters characterizing the background dynamics. A primary quantity of interest is the deceleration parameter, defined as $ q=-\frac{\dot{H}}{H^2}-1$. which directly quantifies the acceleration of the cosmic expansion. Positive values of $q$ correspond to a decelerating Universe, whereas negative values signal an accelerated expansion. 

Another key diagnostic is the effective EoS parameter, $w_{\mathrm{eff}} = \frac{p'}{\rho}$. It is derived directly from the effective pressure and energy density introduced in the field equations beforehand. The effective EoS parameter provides a useful phenomenological description of the dominant cosmic component: $w_{\mathrm{eff}}=0$ corresponds to pressureless matter, $w_{\mathrm{eff}}=1/3$ for radiation era, $w_{\mathrm{eff}}=-1$ for cosmological-constant-like behavior. There are more exotic forms of EoS, like the one with $w_{\mathrm{eff}}<-1$ (phantom fluid), and $-1 < w_{\mathrm{eff}} < -\frac{1}{3}$ as the quintessence phase.

\begin{figure*}
 \centering
  \includegraphics[scale=0.54]{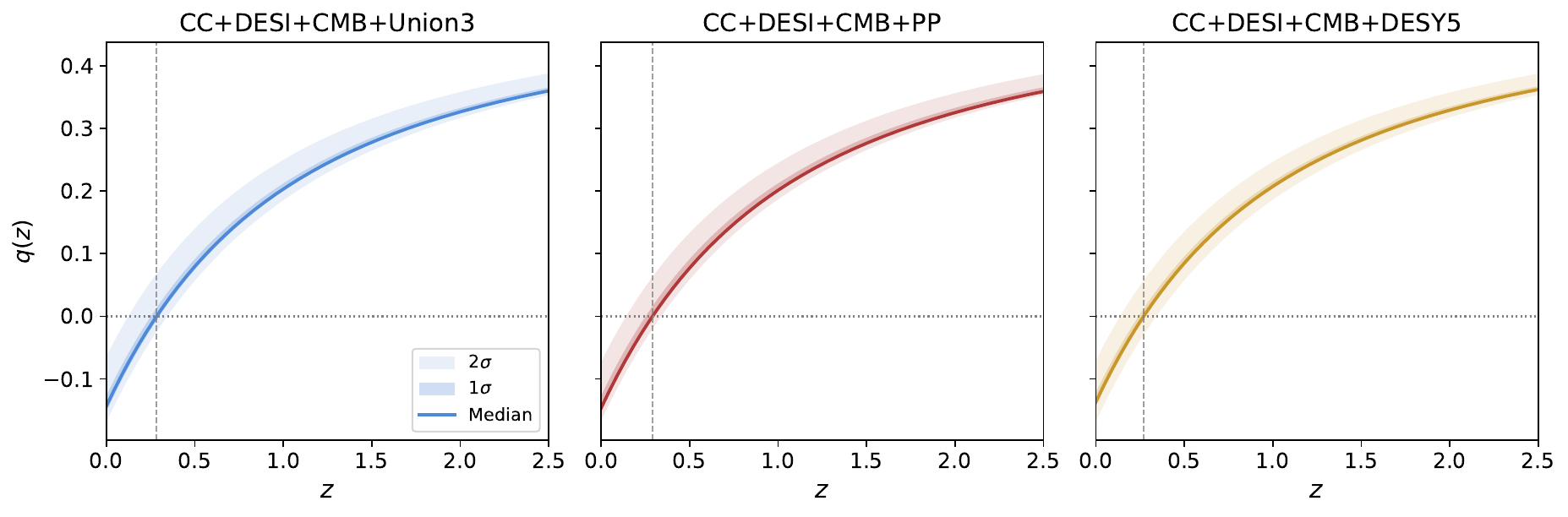}  
 \caption{Deceleration parameter for the viscous GR model reconstructed from observational data. The $1\sigma$ and $2\sigma$ regions are obtained using posterior samples from the MCMC chains.}
 \label{fig:q}
 \end{figure*}

\begin{figure*}
 \centering
 \includegraphics[scale=0.54]{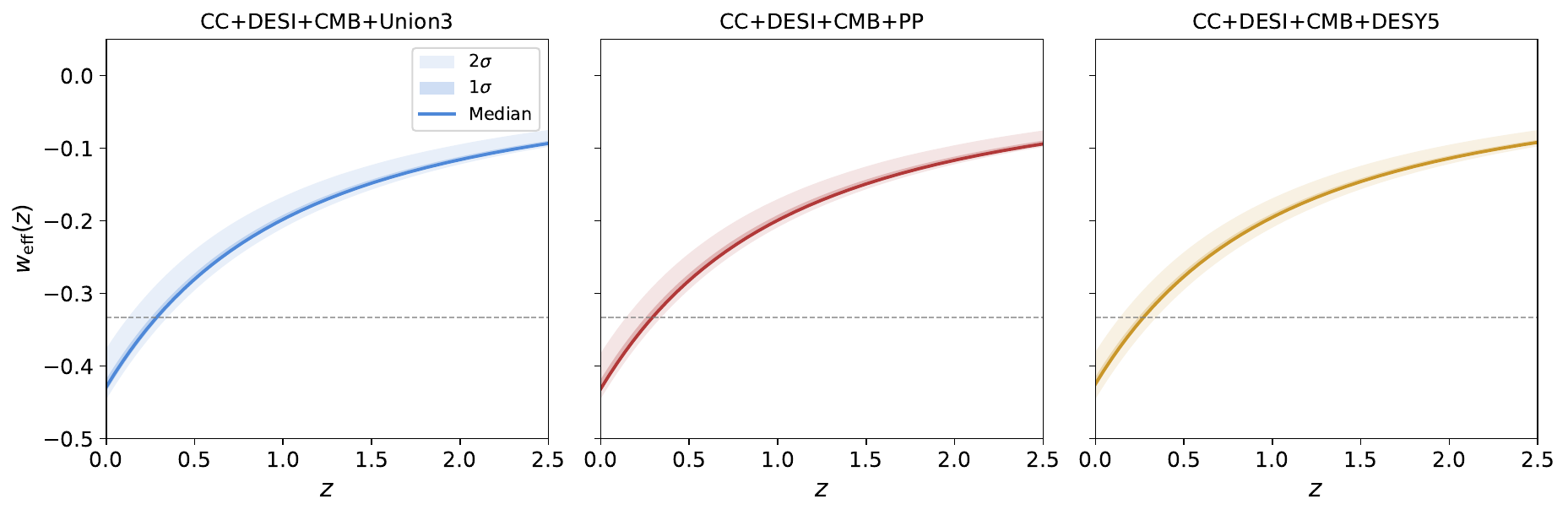}
 \caption{Effective equation of state for the viscous GR model reconstructed from observational data. The confidence regions are obtained using posterior samples from the MCMC chains.}
 \label{fig:w}
 \end{figure*}

\begin{figure*}
 \centering
 \includegraphics[scale=0.48]{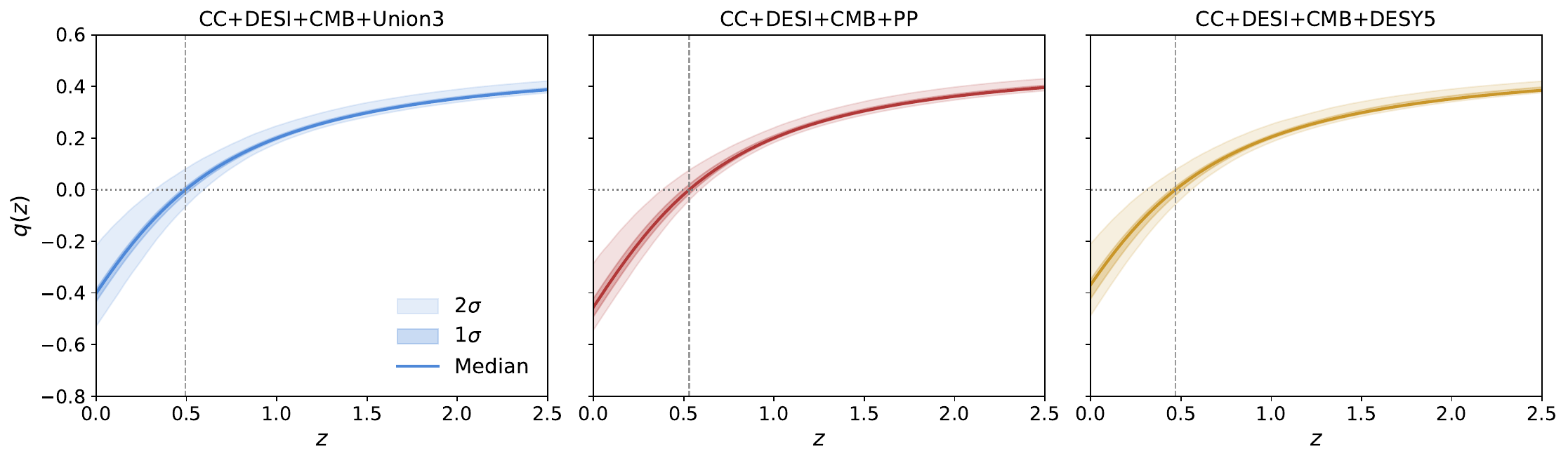}
 \caption{Deceleration parameter for the viscous fluid within the $f(Q)$ framework. The $1\sigma$ and $2\sigma$ confidence regions are reconstructed using posterior samples from the MCMC chains.}
 \label{fig:q2}
 \end{figure*}

\begin{figure*}
 \centering
 \includegraphics[scale=0.48]{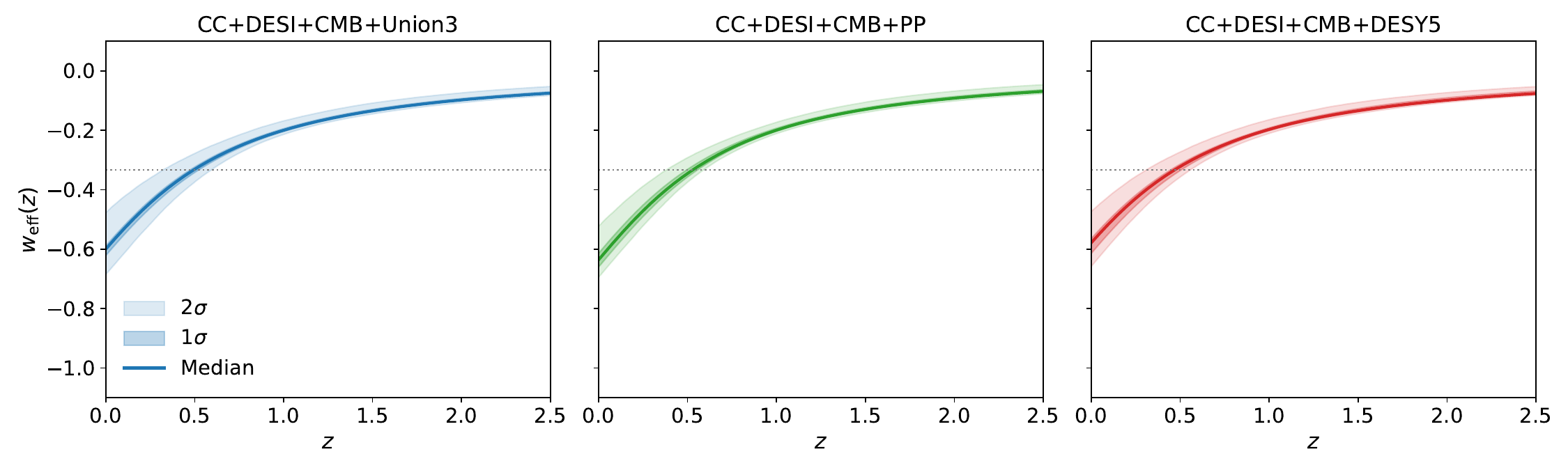}
 \caption{Effective equation of state for the viscous fluid within the $f(Q)$ framework. The $1\sigma$ and $2\sigma$ confidence regions are reconstructed using posterior samples from the MCMC chains.}
 \label{fig:w2}
 \end{figure*}

Figs.~\ref{fig:q} and \ref{fig:w} present the reconstructed evolution of the deceleration parameter, \(q(z)\), and the effective equation-of-state parameter, \(w_{\rm eff}(z)\), for the viscous GR model, while Figs.~\ref{fig:q2} and \ref{fig:w2} show the corresponding results for the exponential \(f(Q)\) model. The reconstructions are obtained from the BASE+PP, BASE+DESY5, and BASE+Union3 dataset combinations. In both models, the reconstructed histories are mutually consistent, indicating that the inferred late-time expansion is only weakly dependent on the choice of supernova compilation. For the viscous GR model, the present-day deceleration parameter is consistently reconstructed as \(q_0\simeq-0.14\), with transition redshifts \(z_t=0.284^{+0.015}_{-0.032}\), \(0.291^{+0.010}_{-0.039}\), and \(0.270^{+0.013}_{-0.028}\) for the BASE+Union3, BASE+PP, and BASE+DESY5 analyses, respectively. The corresponding effective equation of state is remarkably stable, yielding \(w_{\rm eff,0}\simeq-0.43\) across all three dataset combinations. These results indicate a relatively recent onset of cosmic acceleration while maintaining a consistent late-time expansion history.

The exponential \(f(Q)\) model exhibits a systematically different reconstruction. The present-day deceleration parameter lies in the range \(q_0\simeq-(0.37\!-\!0.46)\), with transition redshifts \(z_t\simeq0.47\!-\!0.53\), implying an earlier transition to accelerated expansion than in the viscous GR model. This behaviour is reflected in the effective equation of state, whose present-day value falls in the interval \(w_{\rm eff,0}\simeq-(0.58\!-\!0.64)\). The additional nonmetricity contribution therefore enhances the late-time acceleration while preserving the overall background evolution inferred from the observational data.

 \subsection{Statefinder and Om(z) diagnostics}

The statefinder diagnostics is a widely used practice to differentiate various cosmological models and track their evolution/phase transitions. For such a purpose, the statefinder pair was introduced in the literature \cite{sahni2003statefinder,alam2003exploring}
\begin{eqnarray}
\label{34}
r = \frac{\dddot{a}}{aH^3}, \quad \quad  s= \frac{r-1}{3(q-1/2)}.
\end{eqnarray}
This pair, for the sake of simplicity, could be completely defined in terms of the deceleration parameter
\begin{eqnarray*}
    r(z) &=& q(z)(1+2q(z))+q'(z)(1+z),\\
    \label{32}
    s(z) &=& \frac{r(z)-1}{3(q(z)-1/2)}.
\end{eqnarray*}
In order to perform the aforementioned diagnostics, we need to construct parametric plots within the $r-s$ plane, for which different points and regions correspond to the different states of matter/stages of universe evolution: $\Lambda$CDM corresponds to $(s=0,r=1)$, Chaplygin Gas (CG) corresponds to $(s<0,r>1)$, and quintessence corresponds to $(s>0,r<1)$.

We now turn to a discussion of the well-known and widely used $Om(z)$ diagnostics. This diagnostic was first introduced in \cite{PhysRevD.78.103502,10.1093/mnras/sty755}, where $Om(z)$ is defined as
\begin{equation}
    Om(z)=\frac{E^2(z)-1}{(1+z)^3-1}.
    \label{36}
\end{equation}

\begin{figure*}
 \centering
 \includegraphics[scale=0.47]{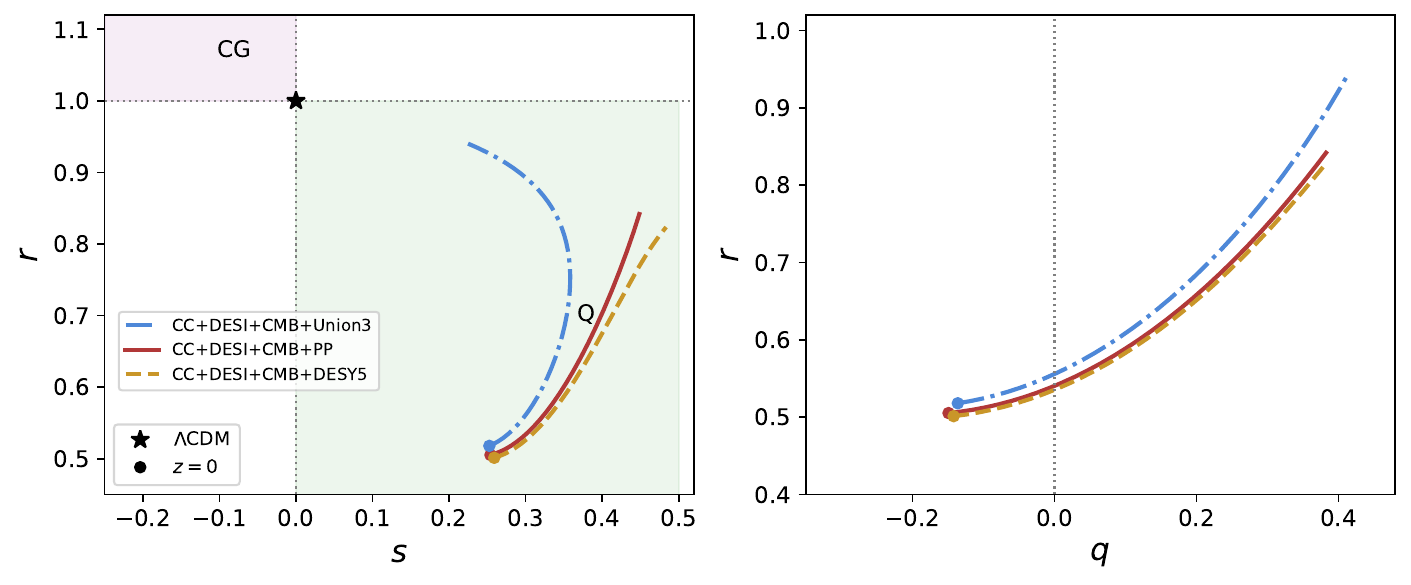}
 \includegraphics[scale=0.42]{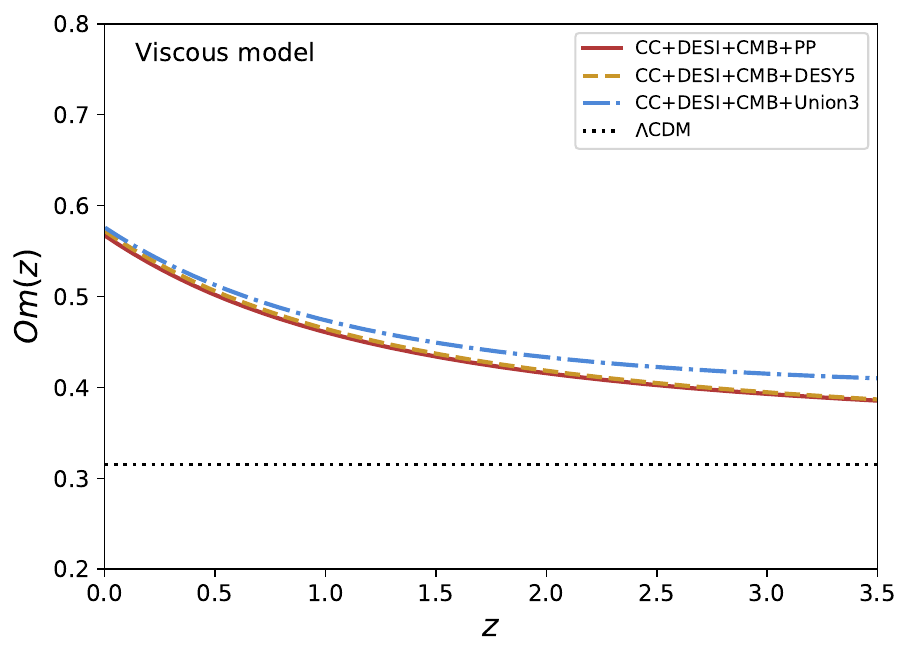} 
 \caption{Parametric trajectories in the $(r,s)$ and $(r,q)$ planes for the viscous GR model. The Chaplygin gas (CG) and quintessence (Q) regions are indicated for comparison. The $Om(z)$ diagnostic is also shown for the considered models, together with the $\Lambda\mathrm{CDM}$ prediction as a reference.}
 \label{fig:SF}
 \end{figure*}

\begin{figure*}
 \centering
 \includegraphics[scale=0.49]{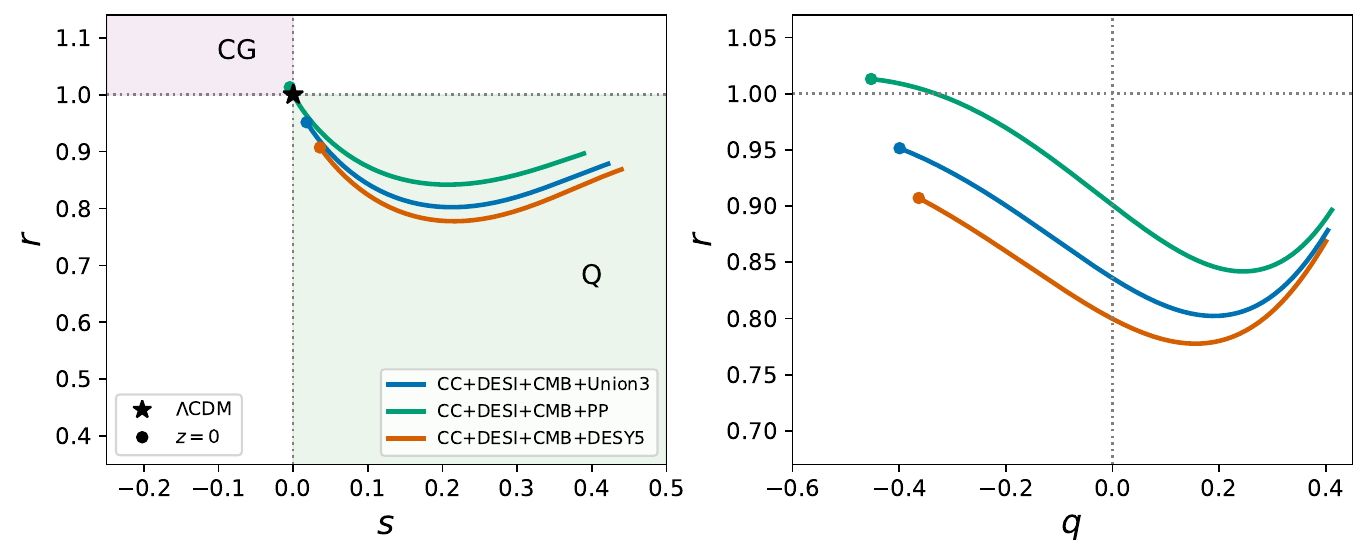}
  \includegraphics[scale=0.42]{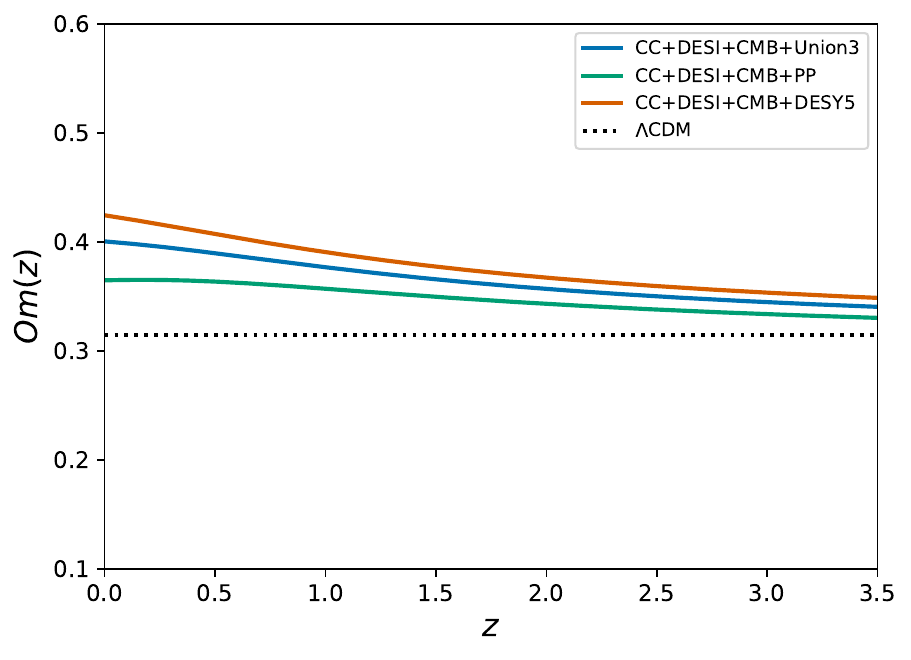}
 \caption{Parametric trajectories of the exponential $f(Q)$ modified gravity model in the $(r,s)$ and $(r,q)$ statefinder planes. The Chaplygin gas (CG) and quintessence (Q) regions are indicated for comparison. The $Om(z)$ diagnostic is also shown for the considered models, together with the $\Lambda\mathrm{CDM}$ prediction as a reference.}
 \label{fig:SF2}
 \end{figure*}

Figs.~\ref{fig:SF} and \ref{fig:SF2} present the statefinder trajectories in the \((r,s)\) and \((r,q)\) planes together with the \(Om(z)\) diagnostic for the viscous GR and exponential \(f(Q)\) models reconstructed from the three observational dataset combinations. In both models, the trajectories originate from the \(\Lambda\)CDM fixed point \((r,s)=(1,0)\) and evolve within the quintessence region while remaining well separated from the Chaplygin gas sector. The three dataset combinations produce nearly overlapping trajectories, indicating that the reconstructed evolution is largely insensitive to the choice of supernova compilation.

Quantitatively, the two models exhibit different behaviours in the geometrical diagnostics. In the \((r,s)\) plane, the exponential \(f(Q)\) trajectories remain closer to the \(\Lambda\)CDM fixed point than those of the viscous GR model, whereas in the \((r,q)\) plane they display a broader evolution before approaching their late-time values. The \(Om(z)\) diagnostic likewise distinguishes the two scenarios: although both models predict a redshift-dependent \(Om(z)\), in contrast to the constant \(\Lambda\)CDM expectation, the exponential \(f(Q)\) model remains systematically closer to the \(\Lambda\)CDM prediction throughout the reconstructed redshift range, while the viscous GR model exhibits a stronger redshift evolution.

\section{Conclusions}
\label{section 7}
Motivated by the extended symmetric teleparallel equivalent of General Relativity, we investigated the cosmological evolution of the Universe in the presence of a viscous fluid within the exponential $f(Q)$ gravity framework. The gravitational sector was described by the exponential form $f(Q)=Qe^{\alpha Q_0/Q}$~\cite{barros2020testing,frusciante2021signatures}, which reduces to General Relativity at high redshifts while allowing departures from the standard cosmological evolution at late times.  The matter sector was modeled by a generalized barotropic fluid with a free equation-of-state parameter $\gamma$, enabling a unified analysis of viscous cosmology within both General Relativity and exponential $f(Q)$ gravity. 

The model parameters were constrained through a Bayesian MCMC analysis using the combined CC+DESI+CMB dataset together with the PP, DESY5, and Union3 Type Ia supernova compilations. The resulting parameter constraints are mutually consistent across the three observational combinations, indicating that the inferred cosmological evolution is robust against the choice of supernova dataset. In all cases, the exponential $f(Q)$ model requires a substantially smaller bulk viscosity than the corresponding viscous GR scenario, demonstrating that the nonmetricity sector contributes significantly to the late-time accelerated expansion.

The reconstructed background evolution shows that both models successfully reproduce the observed expansion history while predicting a smooth transition from an early decelerating Universe to the present accelerated epoch. Nevertheless, the exponential $f(Q)$ model exhibits a systematically earlier transition to acceleration and a more negative present-day effective equation of state than the viscous GR model, indicating a stronger geometric contribution to the late-time dynamics. Throughout the reconstructed redshift range, the effective equation of state remains in the quintessence regime, with no evidence for phantom crossing within the reconstructed confidence intervals.

The phase-space analysis further demonstrates that the cosmological evolution is governed by a stable expanding de Sitter attractor together with an unstable contracting de Sitter solution, while the Minkowski spacetime acts as a semi-stable equilibrium. The reconstructed statefinder trajectories and the $Om(z)$ diagnostic distinguish the two cosmological scenarios while consistently indicating a quintessence-like late-time evolution. In particular, the exponential $f(Q)$ model remains closer to the $\Lambda$CDM prediction in the $Om(z)$ diagnostic, whereas the statefinder analysis reveals distinct higher-order expansion dynamics associated with the nonmetricity sector.

Finally, we performed a statistical comparison with the standard $\Lambda$CDM model using the minimum $\chi^2$, AIC, and BIC. Although both the viscous GR and exponential $f(Q)$ models provide satisfactory fits to the observational data, the information criteria consistently favour the simpler $\Lambda$CDM scenario. Among the two extended models, the exponential $f(Q)$ framework generally provides a better statistical description than the viscous GR model for the BASE+PP and BASE+Union3 datasets, while the viscous GR model is marginally preferred for the BASE+DESY5 combination.

Overall, the present analysis shows that exponential \(f(Q)\) gravity supplemented by a bulk viscous cosmic fluid provides a consistent and observationally viable description of the late-time expansion history. The nonmetricity sector substantially reduces the viscous contribution required to sustain cosmic acceleration and leaves characteristic signatures in the reconstructed background evolution and geometrical diagnostics. While the information criteria favour the minimal \(\Lambda\)CDM model, the exponential \(f(Q)\) scenario remains fully compatible with current background observations and offers a physically motivated extension of General Relativity. Future constraints from structure formation, weak lensing, and forthcoming Stage-IV cosmological surveys will be essential for determining whether these geometric signatures can be distinguished observationally from the standard cosmological model.

\begin{table*}
\centering
\caption{Statistical comparison of viscous GR and viscous exponential $f(Q)$ gravity models relative to $\Lambda$CDM for different combinations of observational datasets.}
\vskip -3mm
\renewcommand{\arraystretch}{1.25}
\begin{tabular}{lcccccc}
\hline\hline
\textbf{Model}
& $\chi^2_{\min}$ 
& $\Delta \chi^2_{\min}$
& AIC
& BIC 
& $\Delta$AIC
& $\Delta$BIC\\
\hline
\multicolumn{7}{c}{\textbf{CC + DESI + CMB + PP}} \\
\hline
$\Lambda$CDM
& 1680.25 & 0 & 1686.25 & 1702.44 & 0 &  0 \\

Viscous
& 1685.66 & 5.41 & 1695.66 & 1722.65 & 9.41 & 20.21 \\

$f(Q)$ 
& 1672.77  & -7.48 & 1684.77 & 1717.16 & -1.48 & 14.72 \\
\hline
\multicolumn{7}{c}{\textbf{CC + DESI + CMB + Union3}} \\
\hline
$\Lambda$CDM & 58.05 & 0 & 66.05 & 74.68 &  0 &  0 \\

Viscous & 62.16 & 4.11 & 72.16 & 82.96 & 6.11 & 8.28 \\

$f(Q)$  & 59.63 & 1.58 & 71.63 & 84.59 & 5.58 & 1.63 \\
\hline
\multicolumn{7}{c}{\textbf{CC + DESI + CMB + DESY5}} \\
\hline
$\Lambda$CDM
& 1684.28 & 0 & 1690.24 &  1712.38 & 0 & 0\\

Viscous
& 1685.47 & 1.19 & 1695.47 & 1723.14 & 5.23 & 10.76 \\

$f(Q)$ 
& 1692.59 & 8.31 & 1704.59 & 1737.80 & 14.35 & 25.42 \\
\hline\hline
\end{tabular}

\label{tab:model_comparison}
\end{table*}

\section*{Data availability} 
There are no new data associated with this article.

\section*{Acknowledgements}
S.A. acknowledges the Japan Society for the Promotion of Science (JSPS) for providing a postdoctoral fellowship during 2024-2026 (JSPS ID No.: P24318). This work of S.A. is also supported by the JSPS KAKENHI grant (Number: 24KF0229). PKS acknowledges Science and Engineering Research Board, Department of Science and Technology, Government of India for financial support to carry out Research project No.: CRG/2022/001847 and IUCAA, Pune, India for providing support through the visiting Associateship program. The authors are very grateful to the honorable referees and the editor for the illuminating suggestions that have significantly improved our work in terms of research quality and presentation.

\bibliography{mybib}

@article{Nair/2016,
    author = "Nair, K. Rajagopalan and Mathew, Titus K.",
    title = "{Bulk viscous Zel\textquoteright{}dovich fluid model and its asymptotic behavior}",
    eprint = "1508.05041",
    archivePrefix = "arXiv",
    primaryClass = "gr-qc",
    doi = "10.1140/epjc/s10052-016-4371-7",
    journal = "Eur. Phys. J. C",
    volume = "76",
    number = "10",
    pages = "519",
    year = "2016"
}

@article{Zel/1962,
    author = "Zel'dovich, Ya. B.",
    title = "{The equation of state at ultrahigh densities and its relativistic limitations}",
    journal = "Zh. Eksp. Teor. Fiz.",
    volume = "41",
    pages = "1609--1615",
    year = "1961",
    doi={}
}

@article{jimenez2020cosmology,
  title={Cosmology in f (Q) geometry},
  author={Jim{\'e}nez, Jose Beltr{\'a}n and Heisenberg, Lavinia and Koivisto, Tomi and Pekar, Simon},
  journal={Physical Review D},
  volume={101},
  number={10},
  pages={103507},
  year={2020},
  publisher={APS}
}

@article{jimenez2018teleparallel,
  title={Teleparallel palatini theories},
  author={Jim{\'e}nez, Jose Beltr{\'a}n and Heisenberg, Lavinia and Koivisto, Tomi S},
  journal={Journal of Cosmology and Astroparticle Physics},
  volume={2018},
  number={08},
  pages={039},
  year={2018},
  publisher={IOP Publishing}
}

@article{Avelino:2008ph,
    author = "Avelino, Arturo and Nucamendi, Ulises",
    title = "{Can a matter-dominated model with constant bulk viscosity drive the accelerated expansion of the universe?}",
    eprint = "0811.3253",
    archivePrefix = "arXiv",
    primaryClass = "gr-qc",
    doi = "10.1088/1475-7516/2009/04/006",
    journal = "JCAP",
    volume = "04",
    pages = "006",
    year = "2009"
}

@article{Mathew:2014gpa,
    author = "Mathew, Titus K. and Aswathy, M. B. and Manoj, M.",
    title = "{Cosmology and thermodynamics of FLRW universe with bulk viscous stiff fluid}",
    eprint = "1406.2089",
    archivePrefix = "arXiv",
    primaryClass = "gr-qc",
    doi = "10.1140/epjc/s10052-014-3188-5",
    journal = "Eur. Phys. J. C",
    volume = "74",
    number = "99",
    pages = "3188",
    year = "2014"
}

@article{Nair:2015bhz,
    author = "Nair, K. Rajagopalan and Mathew, Titus K.",
    title = "{Bulk viscous Zel\textquoteright{}dovich fluid model and its asymptotic behavior}",
    eprint = "1508.05041",
    archivePrefix = "arXiv",
    primaryClass = "gr-qc",
    doi = "10.1140/epjc/s10052-016-4371-7",
    journal = "Eur. Phys. J. C",
    volume = "76",
    number = "10",
    pages = "519",
    year = "2016"
}

@article{Anagnostopoulos:2021ydo,
    author = "Anagnostopoulos, Fotios K. and Basilakos, Spyros and Saridakis, Emmanuel N.",
    title = "{First evidence that non-metricity f(Q) gravity could challenge \ensuremath{\Lambda}CDM}",
    eprint = "2104.15123",
    archivePrefix = "arXiv",
    primaryClass = "gr-qc",
    doi = {10.1016/j.physletb.2021.136634},
    journal = "Phys. Lett. B",
    volume = "822",
    pages = "136634",
    year = "2021"
}

@article{lewis2019getdist,
  title={GetDist: a Python package for analysing Monte Carlo samples},
  author={Lewis, Antony},
  journal={arXiv preprint arXiv:1910.13970},
  year={2019}
}

@article{capozziello2022model,
  title={Model-independent reconstruction of f (Q) non-metric gravity},
  author={Capozziello, Salvatore and D'Agostino, Rocco},
  journal={Physics Letters B},
  pages={137229},
  year={2022},
  publisher={Elsevier},
  doi = {10.1016/j.physletb.2022.137229}
}

@article{Kavya:2025vsj,
    author = "Kavya, N. S. and Mishra, Sai Swagat and Sahoo, P. K.",
    title = "{f(Q) gravity as a possible resolution of the H0 and S8 tensions with DESI DR2}",
    doi = "10.1038/s41598-025-23502-0",
    journal = "Sci. Rep.",
    volume = "15",
    number = "1",
    pages = "36504",
    year = "2025"
}

@article{Mishra:2026mnras,
    author = "Mishra, Sai Swagat and Kavya, N. S. and Sahoo, P. K. and Bamba, Kazuharu",
    title = "{DESI DR2 Meets Cosmography: A Comparative Study of Pad{\'e}, Chebyshev, and Taylor Expansions}",
    eprint = "",
    archivePrefix = "",
    primaryClass = "",
    doi = "10.1093/mnras/stag197",
    journal = "Monthly Notices of the Royal Astronomical Society",
    volume = "546",
    number = "4",
    pages = "stag197",
    year = "2026"
}

@article{Moresco:2020fbm,
    author = "Moresco, Michele and Jimenez, Raul and Verde, Licia and Cimatti, Andrea and Pozzetti, Lucia",
    title = "{Setting the Stage for Cosmic Chronometers. II. Impact of Stellar Population Synthesis Models Systematics and Full Covariance Matrix}",
    eprint = "2003.07362",
    archivePrefix = "arXiv",
    primaryClass = "astro-ph.GA",
    doi = "10.3847/1538-4357/ab9eb0",
    journal = "Astrophys. J.",
    volume = "898",
    number = "1",
    pages = "82",
    year = "2020"
}

@article{Moresco:2012jh,
    author = "Moresco, M. and others",
    title = "{Improved constraints on the expansion rate of the Universe up to z\textasciitilde{}1.1 from the spectroscopic evolution of cosmic chronometers}",
    eprint = "1201.3609",
    archivePrefix = "arXiv",
    primaryClass = "astro-ph.CO",
    doi = "10.1088/1475-7516/2012/08/006",
    journal = "JCAP",
    volume = "08",
    pages = "006",
    year = "2012"
}

@article{Moresco:2016mzx,
    author = "Moresco, Michele and Pozzetti, Lucia and Cimatti, Andrea and Jimenez, Raul and Maraston, Claudia and Verde, Licia and Thomas, Daniel and Citro, Annalisa and Tojeiro, Rita and Wilkinson, David",
    title = "{A 6\% measurement of the Hubble parameter at $z\sim0.45$: direct evidence of the epoch of cosmic re-acceleration}",
    eprint = "1601.01701",
    archivePrefix = "arXiv",
    primaryClass = "astro-ph.CO",
    doi = "10.1088/1475-7516/2016/05/014",
    journal = "JCAP",
    volume = "05",
    pages = "014",
    year = "2016"
}

@article{Jimenez:2003iv,
    author = "Jimenez, Raul and Verde, Licia and Treu, Tommaso and Stern, Daniel",
    title = "{Constraints on the equation of state of dark energy and the Hubble constant from stellar ages and the CMB}",
    eprint = "astro-ph/0302560",
    archivePrefix = "arXiv",
    doi = "10.1086/376595",
    journal = "Astrophys. J.",
    volume = "593",
    pages = "622--629",
    year = "2003"
}

@article{Moresco:2015cya,
    author = "Moresco, Michele",
    title = "{Raising the bar: new constraints on the Hubble parameter with cosmic chronometers at z \ensuremath{\sim} 2}",
    eprint = "1503.01116",
    archivePrefix = "arXiv",
    primaryClass = "astro-ph.CO",
    doi = "10.1093/mnrasl/slv037",
    journal = "Mon. Not. Roy. Astron. Soc.",
    volume = "450",
    number = "1",
    pages = "L16--L20",
    year = "2015"
}

@article{Mishra:2025rhi,
    author = "Mishra, Sai Swagat and Kavya, N. S. and Sahoo, P. K. and Venkatesha, V.",
    title = "{Impact of Teleparallelism on Addressing Current Tensions and Exploring the GW Cosmology}",
    doi = "10.3847/1538-4357/adabc6",
    journal = "Astrophys. J.",
    volume = "981",
    number = "1",
    pages = "13",
    year = "2025"
}

@article{Mishra:2025vpy,
    author = "Mishra, Sai Swagat and Kavya, N. S. and Sahoo, P. K. and Harko, Tiberiu",
    title = "{Pad{\'e} cosmography and its insights into teleparallel gravity}",
    doi = "10.1093/mnras/staf1492",
    journal = "Mon. Not. Roy. Astron. Soc.",
    volume = "543",
    number = "3",
    pages = "2816--2835",
    year = "2025"
}

@article{foremanmackey2013emcee,
  title={emcee: The MCMC hammer},
  author={Foreman-Mackey, Daniel and Hogg, David W and Lang, Dustin and Goodman, Jonathan},
  journal={Publ. Astron. Soc. Pac.},
  volume={125},
  number={925},
  pages={306--312},
  year={2013}
}

@article{Yu_2018,
	doi = {10.3847/1538-4357/aab0a2},
	url = {https://doi.org/10.3847/1538-4357/aab0a2},
	year = 2018,
	month = {mar},
	publisher = {American Astronomical Society},
	volume = {856},
	number = {1},
	pages = {3},
	author = {Hai Yu and Bharat Ratra and Fa-Yin Wang},
	title = {Hubble Parameter and Baryon Acoustic Oscillation Measurement Constraints on the Hubble Constant, the Deviation from the Spatially Flat $\Lambda$CDM Model, the Deceleration{\textendash}Acceleration Transition Redshift, and Spatial Curvature},
	journal = {The Astrophysical Journal}
}

@ARTICLE{2015MNRAS.450L..16M,
       author = {{Moresco}, M.},
        title = "{Raising the bar: new constraints on the Hubble parameter with cosmic chronometers at z \raisebox{-0.5ex}\textasciitilde 2.}",
      journal = {Monthly Notices of the Royal Astronomical Societ},
         year = 2015,
        month = jun,
       volume = {450},
        pages = {L16-L20},
          doi = {10.1093/mnrasl/slv037},
archivePrefix = {arXiv},
       eprint = {1503.01116},
 primaryClass = {astro-ph.CO},
       adsurl = {https://ui.adsabs.harvard.edu/abs/2015MNRAS.450L..16M}
}

@article{Planck:2018vyg,
    author = "Aghanim, N. and others",
    collaboration = "Planck",
    title = "{Planck 2018 results. VI. Cosmological parameters}",
    eprint = "1807.06209",
    archivePrefix = "arXiv",
    primaryClass = "astro-ph.CO",
    doi = "10.1051/0004-6361/201833910",
    journal = "Astron. Astrophys.",
    volume = "641",
    pages = "A6",
    year = "2020",
    note = "[Erratum: Astron.Astrophys. 652, C4 (2021)]"
}

@article{sahni2003statefinder,
  title={Statefinder-a new geometrical diagnostic of dark energy},
  author={Sahni, Varun and Saini, Tarun Deep and Starobinsky, Alexei A and Alam, Ujjaini},
  journal={Journal of Experimental and Theoretical Physics Letters},
  volume={77},
  number={5},
  pages={201--206},
  year={2003},
  publisher={Springer},
  doi={https://doi.org/10.1134/1.1574831}
}

@article{alam2003exploring,
  title={Exploring the expanding universe and dark energy using the Statefinder diagnostic},
  author={Alam, Ujjaini and Sahni, Varun and Deep Saini, Tarun and Starobinsky, AA},
  journal={Monthly Notices of the Royal Astronomical Society},
  volume={344},
  number={4},
  pages={1057--1074},
  year={2003},
  publisher={Blackwell Science Ltd Oxford, UK},
  doi={10.1046/j.1365-8711.2003.06871.x}
}

@article{10.1093/mnras/sty755,
    author = {Pan, Supriya and Mukherjee, Ankan and Banerjee, Narayan},
    title = "{Astronomical bounds on a cosmological model allowing a general interaction in the dark sector}",
    journal = {Monthly Notices of the Royal Astronomical Society},
    volume = {477},
    number = {1},
    pages = {1189-1205},
    year = {2018},
    month = {03},
    issn = {0035-8711},
    doi = {10.1093/mnras/sty755},
    url = {https://doi.org/10.1093/mnras/sty755},
    eprint = {https://academic.oup.com/mnras/article-pdf/477/1/1189/24703456/sty755.pdf},
}

@article{PhysRevD.78.103502,
  title = {Two new diagnostics of dark energy},
  author = {Sahni, Varun and Shafieloo, Arman and Starobinsky, Alexei A.},
  journal = {Phys. Rev. D},
  volume = {78},
  issue = {10},
  pages = {103502},
  numpages = {11},
  year = {2008},
  month = {Nov},
  publisher = {American Physical Society},
  doi = {10.1103/PhysRevD.78.103502},
  url = {https://link.aps.org/doi/10.1103/PhysRevD.78.103502}
}

@article{riess1998observational,
  title={Observational evidence from supernovae for an accelerating universe and a cosmological constant},
  author={Riess, Adam G and Filippenko, Alexei V and Challis, Peter and Clocchiatti, Alejandro and Diercks, Alan and Garnavich, Peter M and Gilliland, Ron L and Hogan, Craig J and Jha, Saurabh and Kirshner, Robert P and others},
  journal={The Astronomical Journal},
  volume={116},
  number={3},
  pages={1009},
  year={1998},
  publisher={IOP Publishing},
  doi={10.1086/300499}
}

@article{perlmutter1999constraining,
  title={Constraining dark energy with type Ia supernovae and large-scale structure},
  author={Perlmutter, Saul and Turner, Michael S and White, Martin},
  journal={Physical Review Letters},
  volume={83},
  number={4},
  pages={670},
  year={1999},
  publisher={APS},
  doi={10.1103/PhysRevLett.83.670}
}

@article{seo2003probing,
  title={Probing dark energy with baryonic acoustic oscillations from future large galaxy redshift surveys},
  author={Seo, Hee-Jong and Eisenstein, Daniel J},
  journal={The Astrophysical Journal},
  volume={598},
  number={2},
  pages={720},
  year={2003},
  publisher={IOP Publishing},
  doi={10.1086/379122}
}

@article{blake2003probing,
  title={Probing dark energy using baryonic oscillations in the galaxy power spectrum as a cosmological ruler},
  author={Blake, Chris and Glazebrook, Karl},
  journal={The Astrophysical Journal},
  volume={594},
  number={2},
  pages={665},
  year={2003},
  publisher={IOP Publishing},
  doi={10.1086/376983}
}

@article{koivisto2006dark,
  title={Dark energy anisotropic stress and large scale structure formation},
  author={Koivisto, Tomi and Mota, David F},
  journal={Physical Review D},
  volume={73},
  number={8},
  pages={083502},
  year={2006},
  publisher={APS},
  doi={10.1103/PhysRevD.73.083502}
}

@article{daniel2008large,
  title={Large scale structure as a probe of gravitational slip},
  author={Daniel, Scott F and Caldwell, Robert R and Cooray, Asantha and Melchiorri, Alessandro},
  journal={Physical Review D},
  volume={77},
  number={10},
  pages={103513},
  year={2008},
  publisher={APS},
  doi={10.1103/PhysRevD.77.103513}
}

@article{cai2016f,
  title={f (T) teleparallel gravity and cosmology},
  author={Cai, Yi-Fu and Capozziello, Salvatore and De Laurentis, Mariafelicia and Saridakis, Emmanuel N},
  journal={Reports on Progress in Physics},
  volume={79},
  number={10},
  pages={106901},
  year={2016},
  publisher={IOP Publishing},
  doi={10.1088/0034-4885/79/10/106901}
}

@article{linder2010einstein,
  title={Einstein's Other Gravity and the Acceleration of the Universe},
  author={Linder, Eric V},
  journal={Physical Review D},
  volume={81},
  number={12},
  pages={127301},
  year={2010},
  publisher={APS},
  doi={10.1103/PhysRevD.81.127301}
}

@article{capozziello2008cosmography,
  title={Cosmography of f (R) gravity},
  author={Capozziello, S and Cardone, VF and Salzano, V},
  journal={Physical Review D},
  volume={78},
  number={6},
  pages={063504},
  year={2008},
  publisher={APS},
  doi={10.1103/PhysRevD.78.063504}
}

@article{sotiriou2010f,
  title={f (R) theories of gravity},
  author={Sotiriou, Thomas P and Faraoni, Valerio},
  journal={Reviews of Modern Physics},
  volume={82},
  number={1},
  pages={451},
  year={2010},
  publisher={APS},
  doi={10.1103/RevModPhys.82.451}
}

@article{de2010f,
  title={f (R) theories},
  author={De Felice, Antonio and Tsujikawa, Shinji},
  journal={Living Reviews in Relativity},
  volume={13},
  number={1},
  pages={1--161},
  year={2010},
  publisher={Springer},
  doi={10.12942/lrr-2010-3}
}

@article{harko2011f,
  title={f (R, T) gravity},
  author={Harko, Tiberiu and Lobo, Francisco SN and Nojiri, Shin’ichi and Odintsov, Sergei D},
  journal={Physical Review D},
  volume={84},
  number={2},
  pages={024020},
  year={2011},
  publisher={APS},
  doi={10.1103/PhysRevD.84.024020}
}

@article{shabani2014cosmological,
  title={Cosmological and solar system consequences of f (R, T) gravity models},
  author={Shabani, Hamid and Farhoudi, Mehrdad},
  journal={Physical Review D},
  volume={90},
  number={4},
  pages={044031},
  year={2014},
  publisher={APS},
  doi={10.1103/PhysRevD.90.044031}
}

@article{Hehl:1976kj,
    author = "Hehl, F. W. and Von Der Heyde, P. and Kerlick, G. D. and Nester, J. M.",
    title = "{General Relativity with Spin and Torsion: Foundations and Prospects}",
    doi = "10.1103/RevModPhys.48.393",
    journal = "Rev. Mod. Phys.",
    volume = "48",
    pages = "393--416",
    year = "1976"
}

@article{Gadbail:2024als,
    author = "Gadbail, Gaurav N. and Arora, Simran and Channuie, Phongpichit and Sahoo, P. K.",
    title = "{Cosmological Dynamics of Interacting Dark Energy and Dark Matter in f(Q) Gravity}",
    eprint = "2406.02026",
    archivePrefix = "arXiv",
    primaryClass = "gr-qc",
    doi = "10.1002/prop.202400205",
    journal = "Fortsch. Phys.",
    volume = "73",
    number = "5",
    pages = "2400205",
    year = "2025"
}

@article{Israel:1979wp,
    author = "Israel, W. and Stewart, J. M.",
    title = "{Transient relativistic thermodynamics and kinetic theory}",
    doi = "10.1016/0003-4916(79)90130-1",
    journal = "Annals Phys.",
    volume = "118",
    pages = "341--372",
    year = "1979"
}

@article{Weinberg:1988cp,
    author = "Weinberg, Steven",
    editor = "Hsu, Jong-Ping and Fine, D.",
    title = "{The Cosmological Constant Problem}",
    reportNumber = "UTTG-12-88",
    doi = "10.1103/RevModPhys.61.1",
    journal = "Rev. Mod. Phys.",
    volume = "61",
    pages = "1--23",
    year = "1989"
}

@article{Zlatev:1998tr,
    author = "Zlatev, Ivaylo and Wang, Li-Min and Steinhardt, Paul J.",
    title = "{Quintessence, cosmic coincidence, and the cosmological constant}",
    eprint = "astro-ph/9807002",
    archivePrefix = "arXiv",
    doi = "10.1103/PhysRevLett.82.896",
    journal = "Phys. Rev. Lett.",
    volume = "82",
    pages = "896--899",
    year = "1999"
}

@article{Avelino:2025lki,
    author = "Avelino, P. P. and Gomes, A. R. and Tamayo, D. A.",
    title = "{Note on bulk viscosity as an alternative to dark energy}",
    eprint = "2512.15633",
    archivePrefix = "arXiv",
    primaryClass = "astro-ph.CO",
    doi = "10.1103/2kf6-413m",
    journal = "Phys. Rev. D",
    volume = "112",
    number = "12",
    pages = "123531",
    year = "2025"
}

@article{Palma:2025qge,
    author = "Palma, Guillermo and Gomez, Gabriel",
    title = "{Non-linear causal bulk viscosity in unified dark matter cosmologies}",
    eprint = "2510.11900",
    archivePrefix = "arXiv",
    primaryClass = "gr-qc",
    doi = "10.1140/epjc/s10052-025-15213-7",
    journal = "Eur. Phys. J. C",
    volume = "85",
    number = "12",
    pages = "1486",
    year = "2025"
}

@article{Villalobos:2025mdk,
    author = "Villalobos, R. Noem{\'\i} and V{\'a}squez, Yerko and Cruz, Norman and L{\'o}pez-Caraballo, Carlos H.",
    title = "{Bulk viscous cosmological models with cosmological constant: Observational constraints}",
    eprint = "2508.11614",
    archivePrefix = "arXiv",
    primaryClass = "astro-ph.CO",
    month = "8",
    doi = "",
    journal = "",
    volume = "",
    number = "",
    pages = "",
    year = "2025"
}

@article{zhao2022covariant,
  title={Covariant formulation of f(Q) theory},
  author={Zhao, Dehao},
  journal={The European Physical Journal C},
  volume={82},
  number={4},
  pages={1--12},
  year={2022},
  publisher={Springer},
  doi={10.1140/epjc/s10052-022-10266-4}
}

@article{Wang:2025bkk,
    author = "Wang, Deng and Mota, David",
    title = "{Did DESI DR2 truly reveal dynamical dark energy?}",
    eprint = "2504.15222",
    archivePrefix = "arXiv",
    primaryClass = "astro-ph.CO",
doi = "",
    journal = "",
    volume = "",
    pages = "",
    month = "4",
    year = "2025"
}

@article{Mishra:2025goj,
    author = "Mishra, Swagat S. and Matthewson, William L. and Sahni, Varun and Shafieloo, Arman and Shtanov, Yuri",
    title = "{Braneworld dark energy in light of DESI~DR2}",
    eprint = "2507.07193",
    archivePrefix = "arXiv",
    primaryClass = "astro-ph.CO",
    doi = "10.1088/1475-7516/2025/11/018",
    journal = "JCAP",
    volume = "11",
    pages = "018",
    year = "2025"
}

@article{PhysRevD.106.043509,
  title = {FLRW solutions in $f(Q)$ theory: The effect of using different connections},
  author = {Dimakis, N. and Paliathanasis, A. and Roumeliotis, M. and Christodoulakis, T.},
  journal = {Physical Review DD},
  volume = {106},
  issue = {4},
  pages = {043509},
  numpages = {17},
  year = {2022},
  month = {Aug},
  publisher = {American Physical Society},
  doi = {10.1103/PhysRevD.106.043509},
  url = {https://link.aps.org/doi/10.1103/PhysRevD.106.043509}
}

@article{d2022black,
  title={Black holes in f (Q) gravity},
  author={D'Ambrosio, Fabio and Fell, Shaun DB and Heisenberg, Lavinia and Kuhn, Simon},
  journal={Physical Review D},
  volume={105},
  number={2},
  pages={024042},
  year={2022},
  publisher={APS},
  doi={10.1103/PhysRevD.105.024042}
}

@article{mandal2020cosmography,
  title={Cosmography in f (Q) gravity},
  author={Mandal, Sanjay and Wang, Deng and Sahoo, PK},
  journal={Physical Review D},
  volume={102},
  number={12},
  pages={124029},
  year={2020},
  publisher={APS},
  doi={10.1103/PhysRevD.102.124029}
}

@article{Arora:2025msq,
    author = "Arora, Simran and De Felice, Antonio and Mukohyama, Shinji",
    title = "{Dynamical dark energy parametrizations in VCDM}",
    eprint = "2508.03784",
    archivePrefix = "arXiv",
    primaryClass = "gr-qc",
    doi = "10.1103/l5bx-snl3",
    journal = "Phys. Rev. D",
    volume = "112",
    number = "12",
    pages = "123518",
    year = "2025"
}

@article{DESI:2025wyn,
    author = "Gu, Gan and others",
    collaboration = "DESI",
    title = "{Dynamical dark energy in light of the DESI DR2 baryonic acoustic oscillations measurements}",
    eprint = "2504.06118",
    archivePrefix = "arXiv",
    primaryClass = "astro-ph.CO",
    reportNumber = "FERMILAB-PUB-25-0235-PPD",
    doi = "10.1038/s41550-025-02669-6",
    journal = "Nature Astron.",
    volume = "9",
    number = "12",
    pages = "1879--1889",
    year = "2025",
    note = "[Erratum: Nature Astron. 9, 1898 (2025)]"
}

@article{Khyllep:2022spx,
    author = "Khyllep, Wompherdeiki and Dutta, Jibitesh and Saridakis, Emmanuel N. and Yesmakhanova, Kuralay",
    title = "{Cosmology in f(Q) gravity: A unified dynamical systems analysis of the background and perturbations}",
    eprint = "2207.02610",
    archivePrefix = "arXiv",
    primaryClass = "gr-qc",
    doi = "10.1103/PhysRevD.107.044022",
    journal = "Phys. Rev. D",
    volume = "107",
    number = "4",
    pages = "044022",
    year = "2023"
}

@article{Sokoliuk:2023ccw,
    author = "Sokoliuk, Oleksii and Arora, Simran and Praharaj, Subhrat and Baransky, Alexander and Sahoo, P. K.",
    title = "{On the impact of f(Q) gravity on the large scale structure}",
    eprint = "2303.17341",
    archivePrefix = "arXiv",
    primaryClass = "astro-ph.CO",
    doi = "10.1093/mnras/stad968",
    journal = "Mon. Not. Roy. Astron. Soc.",
    volume = "522",
    number = "1",
    pages = "252--267",
    year = "2023"
}

@article{Anagnostopoulos:2022gej,
    author = "Anagnostopoulos, Fotios K. and Gakis, Viktor and Saridakis, Emmanuel N. and Basilakos, Spyros",
    title = "{New models and big bang nucleosynthesis constraints in f(Q) gravity}",
    eprint = "2205.11445",
    archivePrefix = "arXiv",
    primaryClass = "gr-qc",
    doi = "10.1140/epjc/s10052-023-11190-x",
    journal = "Eur. Phys. J. C",
    volume = "83",
    number = "1",
    pages = "58",
    year = "2023"
}

@article{Arora:2022mlo,
    author = "Arora, Simran and Sahoo, Pradyumn Kumar",
    title = "{Crossing Phantom Divide in f(Q)$f(Q)$ Gravity}",
    eprint = "2206.05110",
    archivePrefix = "arXiv",
    primaryClass = "gr-qc",
    doi = "10.1002/andp.202200233",
    journal = "Annalen Phys.",
    volume = "534",
    number = "8",
    pages = "2200233",
    year = "2022"
}

@article{albuquerque2022designer,
  title={A designer approach to f (Q) gravity and cosmological implications},
  author={Albuquerque, In{\^e}s S and Frusciante, Noemi},
  journal={Physics of the Dark Universe},
  volume={35},
  pages={100980},
  year={2022},
  publisher={Elsevier},
  doi={10.1016/j.dark.2022.100980}
}

@article{lymperis2022late,
  title={Late-time cosmology with phantom dark-energy in f (Q) gravity},
  author={Lymperis, Andreas},
  journal={Journal of Cosmology and Astroparticle Physics},
  volume={2022},
  number={11},
  pages={018},
  year={2022},
  publisher={IOP Publishing},
  doi={10.1088/1475-7516/2022/11/018}
}

@article{frusciante2021signatures,
  title={Signatures of f (Q) gravity in cosmology},
  author={Frusciante, Noemi},
  journal={Physical Review D},
  volume={103},
  number={4},
  pages={044021},
  year={2021},
  publisher={APS},
  doi={10.1103/PhysRevD.103.044021}
}

@article{atayde2021can,
  title={Can f (Q) gravity challenge $\Lambda$ CDM?},
  author={Atayde, Lu{\'\i}s and Frusciante, Noemi},
  journal={Physical Review D},
  volume={104},
  number={6},
  pages={064052},
  year={2021},
  publisher={APS},
  doi={10.1103/PhysRevD.104.064052}
}

@article{lazkoz2019observational,
  title={Observational constraints of f (Q) gravity},
  author={Lazkoz, Ruth and Lobo, Francisco SN and Ortiz-Ba{\~n}os, Mar{\'\i}a and Salzano, Vincenzo},
  journal={Physical Review D},
  volume={100},
  number={10},
  pages={104027},
  year={2019},
  publisher={APS},
  doi={10.1103/PhysRevD.100.104027}
}

@article{d2022forecasting,
    author = "D'Agostino, Rocco and Nunes, Rafael C.",
    title = "{Forecasting constraints on deviations from general relativity in f(Q) gravity with standard sirens}",
    eprint = "2210.11935",
    archivePrefix = "arXiv",
    primaryClass = "gr-qc",
    reportNumber = "ET-0236A-22",
    doi = "10.1103/PhysRevD.106.124053",
    journal = "Phys. Rev. D",
    volume = "106",
    number = "12",
    pages = "124053",
    year = "2022"
}

@article{ayuso2021observational,
  title={Observational constraints on cosmological solutions of f (Q) theories},
  author={Ayuso, Ismael and Lazkoz, Ruth and Salzano, Vincenzo},
  journal={Physical Review D},
  volume={103},
  number={6},
  pages={063505},
  year={2021},
  publisher={APS},
  doi={10.1103/PhysRevD.103.063505}
}

@article{barros2020testing,
  title={Testing F (Q) gravity with redshift space distortions},
  author={Barros, Bruno J and Barreiro, Tiago and Koivisto, Tomi and Nunes, Nelson J},
  journal={Physics of the Dark Universe},
  volume={30},
  pages={100616},
  year={2020},
  publisher={Elsevier},
  doi={10.1016/j.dark.2020.100616}
}

@article{anagnostopoulos2021first,
  title={First evidence that non-metricity f (Q) gravity could challenge $\Lambda$CDM},
  author={Anagnostopoulos, Fotios K and Basilakos, Spyros and Saridakis, Emmanuel N},
  journal={Physics Letters B},
  volume={822},
  pages={136634},
  year={2021},
  publisher={Elsevier}
}

@article{dialektopoulos2019noether,
  title={Noether symmetries in symmetric teleparallel cosmology},
  author={Dialektopoulos, Konstantinos F and Koivisto, Tomi S and Capozziello, Salvatore},
  journal={The European Physical Journal C},
  volume={79},
  number={7},
  pages={1--12},
  year={2019},
  publisher={Springer}
}

@article{jimenez2018coincident,
  title={Coincident general relativity},
  author={Jim{\'e}nez, Jose Beltr{\'a}n and Heisenberg, Lavinia and Koivisto, Tomi},
  journal={Physical Review D},
  volume={98},
  number={4},
  pages={044048},
  year={2018},
  publisher={APS}
}

@article{Awad:2017yod,
    author = "Awad, A. and El Hanafy, W. and Nashed, G. G. L. and Saridakis, Emmanuel N.",
    title = "{Phase Portraits of general f(T) Cosmology}",
    eprint = "1710.10194",
    archivePrefix = "arXiv",
    primaryClass = "gr-qc",
    doi = "10.1088/1475-7516/2018/02/052",
    journal = "JCAP",
    volume = "02",
    pages = "052",
    year = "2018"
}

@article{singh2014friedmann,
  title={Friedmann model with viscous cosmology in modified $f(R, T)$ gravity theory},
  author={Singh, CP and Kumar, Pankaj},
  journal={The European Physical Journal C},
  volume={74},
  number={10},
  pages={1--11},
  year={2014},
  publisher={Springer},
  doi={10.1140/epjc/s10052-014-3070-5}
}

@article{solanki2021cosmic,
  title={Cosmic acceleration with bulk viscosity in modified f (Q) gravity},
  author={Solanki, Raja and Pacif, SKJ and Parida, Abhishek and Sahoo, PK},
  journal={Physics of the Dark Universe},
  volume={32},
  pages={100820},
  year={2021},
  publisher={Elsevier},
  doi={10.1016/j.dark.2021.100820}
}

@article{arora2022bulk,
  title={Bulk viscous matter and the cosmic acceleration of the universe in f (Q, T) gravity},
  author={Arora, Simran and Pacif, SKJ and Parida, Abhishek and Sahoo, PK},
  journal={Journal of High Energy Astrophysics},
  volume={33},
  pages={1--9},
  year={2022},
  publisher={Elsevier},
  doi={10.1016/j.jheap.2021.10.001}
}

@article{brevik2006crossing,
  title={Crossing of the w=-1 barrier in viscous modified gravity},
  author={Brevik, Iver},
  journal={International Journal of Modern Physics D},
  volume={15},
  number={05},
  pages={767--775},
  year={2006},
  publisher={World Scientific},
  doi={10.1142/S0218271806008528}
}

@article{mathew2014cosmology,
  title={Cosmology and thermodynamics of FLRW universe with bulk viscous stiff fluid},
  author={Mathew, Titus K and Aswathy, MB and Manoj, M},
  journal={The European Physical Journal C},
  volume={74},
  number={12},
  pages={1--10},
  year={2014},
  publisher={Springer},
  doi={10.1140/epjc/s10052-014-3188-5}
}

@article{eckart1940thermodynamics,
  title={The thermodynamics of irreversible processes. III. Relativistic theory of the simple fluid},
  author={Eckart, Carl},
  journal={Physical review},
  volume={58},
  number={10},
  pages={919},
  year={1940},
  publisher={APS},
  doi={10.1103/PhysRev.58.919}
}

@article{ren2006modified,
  title={Modified equation of state, scalar field, and bulk viscosity in Friedmann universe},
  author={Ren, Jie and Meng, Xin-He},
  journal={Physics Letters B},
  volume={636},
  number={1},
  pages={5--12},
  year={2006},
  publisher={Elsevier},
  doi={10.1016/j.physletb.2006.03.029}
}

@article{DESI:2024mwx,
    author = "Adame, A. G. and others",
    collaboration = "DESI",
    title = "{DESI 2024 VI: cosmological constraints from the measurements of baryon acoustic oscillations}",
    eprint = "2404.03002",
    archivePrefix = "arXiv",
    primaryClass = "astro-ph.CO",
    reportNumber = "FERMILAB-PUB-24-0154-PPD",
    doi = "10.1088/1475-7516/2025/02/021",
    journal = "JCAP",
    volume = "02",
    pages = "021",
    year = "2025"
}

@article{DESI:2025zgx,
    author = "Abdul Karim, M. and others",
    title = "{DESI DR2 Results II: Measurements of Baryon Acoustic Oscillations and Cosmological Constraints}",
    eprint = "2503.14738",
    archivePrefix = "arXiv",
    primaryClass = "astro-ph.CO",
    reportNumber = "FERMILAB-PUB-25-0169-PPD",
    month = "3",
    year = "2025",
doi = "",
journal = ""
}

@article{Moon:2023jgl,
	author = "Moon, Jeongin and others",
	title = "{First detection of the BAO signal from early DESI data}",
	eprint = "2304.08427",
	archivePrefix = "arXiv",
	primaryClass = "astro-ph.CO",
	reportNumber = "FERMILAB-PUB-23-199-PPD",
	doi = "10.1093/mnras/stad2618",
	journal = "Mon. Not. Roy. Astron. Soc.",
	volume = "525",
	number = "4",
	pages = "5406--5422",
	year = "2023"
}

@article{eBOSS:2020yzd,
	author = "Alam, Shadab and others",
	collaboration = "eBOSS",
	title = "{Completed SDSS-IV extended Baryon Oscillation Spectroscopic Survey: Cosmological implications from two decades of spectroscopic surveys at the Apache Point Observatory}",
	eprint = "2007.08991",
	archivePrefix = "arXiv",
	primaryClass = "astro-ph.CO",
	doi = "10.1103/PhysRevD.103.083533",
	journal = "Phys. Rev. D",
	volume = "103",
	number = "8",
	pages = "083533",
	year = "2021"
}

@article{Rubin:2023jdq,
    author = "Rubin, David and others",
    title = "{Union Through UNITY: Cosmology with 2,000 SNe Using a Unified Bayesian Framework}",
    eprint = "2311.12098",
    archivePrefix = "arXiv",
    primaryClass = "astro-ph.CO",
    doi = "10.3847/1538-4357/adc0a5",
    journal = "Astrophys. J.",
    volume = "986",
    number = "2",
    pages = "231",
    year = "2025"
}

@article{Kolhatkar:2024oyy,
    author = "Kolhatkar, Ameya and Mishra, Sai Swagat and Sahoo, P. K.",
    title = "{Investigating early and late-time epochs in f(Q) gravity}",
    eprint = "2409.01538",
    archivePrefix = "arXiv",
    primaryClass = "gr-qc",
    doi = "10.1140/epjc/s10052-024-13237-z",
    journal = "Eur. Phys. J. C",
    volume = "84",
    number = "9",
    pages = "888",
    year = "2024"
}

@article{Brout_2022,
	doi = {10.3847/1538-4357/ac8e04},
	url = {https://dx.doi.org/10.3847/1538-4357/ac8e04},
	year = {2022},
	month = {oct},
	publisher = {The American Astronomical Society},
	volume = {938},
	number = {2},
	pages = {110},
	author = "Dillon Brout et al.",
	title = {The Pantheon+ Analysis: Cosmological Constraints},
	journal = {The Astrophysical Journal},
}

@article{DES:2024jxu,
    author = "Abbott, T. M. C. and others",
    collaboration = "DES",
    title = "{The Dark Energy Survey: Cosmology Results with {\ensuremath{\sim}}1500 New High-redshift Type Ia Supernovae Using the Full 5 yr Data Set}",
    eprint = "2401.02929",
    archivePrefix = "arXiv",
    primaryClass = "astro-ph.CO",
    reportNumber = "FERMILAB-PUB-23-0821-PPD, DES-2023-805",
    doi = "10.3847/2041-8213/ad6f9f",
    journal = "Astrophys. J. Lett.",
    volume = "973",
    number = "1",
    pages = "L14",
    year = "2024"
}

@article{DES:2024hip,
    author = "Vincenzi, M. and others",
    collaboration = "DES",
    title = "{The Dark Energy Survey Supernova Program: Cosmological Analysis and Systematic Uncertainties}",
    eprint = "2401.02945",
    archivePrefix = "arXiv",
    primaryClass = "astro-ph.CO",
    reportNumber = "FERMILAB-PUB-23-693-PPD",
    doi = "10.3847/1538-4357/ad5e6c",
    journal = "Astrophys. J.",
    volume = "975",
    number = "1",
    pages = "86",
    year = "2024"
}

@article{Hu:1995en,
	author = "Hu, Wayne and Sugiyama, Naoshi",
	title = "{Small scale cosmological perturbations: An Analytic approach}",
	eprint = "astro-ph/9510117",
	archivePrefix = "arXiv",
	reportNumber = "IASSNS-AST-95-42, CFPA-TH-95-18, UTAP-212",
	doi = "10.1086/177989",
	journal = "Astrophys. J.",
	volume = "471",
	pages = "542--570",
	year = "1996"
}

@article{Chen:2018dbv,
	author = "Chen, Lu and Huang, Qing-Guo and Wang, Ke",
	title = "{Distance Priors from Planck Final Release}",
	eprint = "1808.05724",
	archivePrefix = "arXiv",
	primaryClass = "astro-ph.CO",
	doi = "10.1088/1475-7516/2019/02/028",
	journal = "JCAP",
	volume = "02",
	pages = "028",
	year = "2019"
}

@article{Arendse:2019hev,
	author = "Arendse, Nikki and others",
	title = "{Cosmic dissonance: are new physics or systematics behind a short sound horizon?}",
	eprint = "1909.07986",
	archivePrefix = "arXiv",
	primaryClass = "astro-ph.CO",
	doi = "10.1051/0004-6361/201936720",
	journal = "Astron. Astrophys.",
	volume = "639",
	pages = "A57",
	year = "2020"
}

@article{Cai:2014ela,
    author = "Cai, Rong-Gen and Guo, Zong-Kuan and Tang, Bo",
    title = "{Updated reduced CMB data and constraints on cosmological parameters}",
    eprint = "1409.0223",
    archivePrefix = "arXiv",
    primaryClass = "astro-ph.CO",
    doi = "10.1142/S0218271815500716",
    journal = "Int. J. Mod. Phys. D",
    volume = "24",
    number = "10",
    pages = "1550071",
    year = "2015"
}

@article{Liu:2024fjy,
    author = "Liu, Guanlin and Wang, Yu and Zhao, Wen",
    title = "{Testing the consistency of early and late cosmological parameters with BAO and CMB data}",
    eprint = "2401.10571",
    archivePrefix = "arXiv",
    primaryClass = "astro-ph.CO",
    doi = "10.1016/j.physletb.2024.138717",
    journal = "Phys. Lett. B",
    volume = "854",
    pages = "138717",
    year = "2024"
}

@article{Khyllep:2021pcu,
    author = "Khyllep, Wompherdeiki and Paliathanasis, Andronikos and Dutta, Jibitesh",
    title = "{Cosmological solutions and growth index of matter perturbations in $f(Q)$ gravity}",
    eprint = "2103.08372",
    archivePrefix = "arXiv",
    primaryClass = "gr-qc",
    doi = "10.1103/PhysRevD.103.103521",
    journal = "Phys. Rev. D",
    volume = "103",
    number = "10",
    pages = "103521",
    year = "2021"
}

@article{Burnham2004,
author = "Burnham, Kenneth P. and Anderson, David R.",
title = "{Multimodel Inference: Understanding AIC and BIC in Model Selection}",
doi = "10.1177/0049124104268644",
journal = "Sociological Methods \& Research",
volume = "33",
number = "2",
pages = "261--304",
year = "2004"
}

@article{Alestas:2022gcg,
    author = "Alestas, George and Kazantzidis, Lavrentios and Nesseris, Savvas",
    title = "{Machine learning constraints on deviations from general relativity from the large scale structure of the Universe}",
    eprint = "2209.12799",
    archivePrefix = "arXiv",
    primaryClass = "astro-ph.CO",
    reportNumber = "IFT-UAM/CSIC-22-112",
    doi = "10.1103/PhysRevD.106.103519",
    journal = "Phys. Rev. D",
    volume = "106",
    number = "10",
    pages = "103519",
    year = "2022"
}

@article{Macaulay:2013swa,
    author = "Macaulay, Edward and Wehus, Ingunn Kathrine and Eriksen, Hans Kristian",
    title = "{Lower Growth Rate from Recent Redshift Space Distortion Measurements than Expected from Planck}",
    eprint = "1303.6583",
    archivePrefix = "arXiv",
    primaryClass = "astro-ph.CO",
    doi = "10.1103/PhysRevLett.111.161301",
    journal = "Phys. Rev. Lett.",
    volume = "111",
    number = "16",
    pages = "161301",
    year = "2013"
}

@article{Nesseris:2017vor,
    author = "Nesseris, Savvas and Pantazis, George and Perivolaropoulos, Leandros",
    title = "{Tension and constraints on modified gravity parametrizations of $G_{\textrm{eff}}(z)$ from growth rate and Planck data}",
    eprint = "1703.10538",
    archivePrefix = "arXiv",
    primaryClass = "astro-ph.CO",
    reportNumber = "IFT-UAM-CSIC-17-031",
    doi = "10.1103/PhysRevD.96.023542",
    journal = "Phys. Rev. D",
    volume = "96",
    number = "2",
    pages = "023542",
    year = "2017"
}

@article{Elgaroy:2007bv,
    author = "Elgaroy, Oystein and Multamaki, Tuomas",
    title = "{On using the CMB shift parameter in tests of models of dark energy}",
    eprint = "astro-ph/0702343",
    archivePrefix = "arXiv",
    doi = "10.1051/0004-6361:20077292",
    journal = "Astron. Astrophys.",
    volume = "471",
    pages = "65",
    year = "2007"
}

@article{Trotta:2008qt,
    author = "Trotta, Roberto",
    title = "{Bayes in the sky: Bayesian inference and model selection in cosmology}",
    eprint = "0803.4089",
    archivePrefix = "arXiv",
    primaryClass = "astro-ph",
    doi = "10.1080/00107510802066753",
    journal = "Contemp. Phys.",
    volume = "49",
    pages = "71--104",
    year = "2008"
}

@article{Kass:1995loi,
    author = "Kass, Robert E. and Raftery, Adrian E.",
    title = "{Bayes Factors}",
    doi = "10.1080/01621459.1995.10476572",
    journal = "J. Am. Statist. Assoc.",
    volume = "90",
    number = "430",
    pages = "773--795",
    year = "1995"
}

@article{Akaike:1974vps,
    author = "Akaike, H.",
    title = "{A new look at the statistical model identification}",
    doi = "10.1109/TAC.1974.1100705",
    journal = "IEEE Trans. Automatic Control",
    volume = "19",
    number = "6",
    pages = "716--723",
    year = "1974"
}

@article{Liddle:2007fy,
    author = "Liddle, Andrew R",
    title = "{Information criteria for astrophysical model selection}",
    eprint = "astro-ph/0701113",
    archivePrefix = "arXiv",
    doi = "10.1111/j.1745-3933.2007.00306.x",
    journal = "Mon. Not. Roy. Astron. Soc.",
    volume = "377",
    pages = "L74--L78",
    year = "2007"
}

@article{schwarz1978estimating,
  title={Estimating the dimension of a model},
  author={Schwarz, Gideon},
  journal={The annals of statistics},
  pages={461--464},
  year={1978},
  publisher={JSTOR}
}

@article{Arevalo:2016epc,
    author = "Arevalo, Fabiola and Cid, Antonella and Moya, Jorge",
    title = "{AIC and BIC for cosmological interacting scenarios}",
    eprint = "1610.09330",
    archivePrefix = "arXiv",
    primaryClass = "astro-ph.CO",
    doi = "10.1140/epjc/s10052-017-5128-7",
    journal = "Eur. Phys. J. C",
    volume = "77",
    number = "8",
    pages = "565",
    year = "2017"
}
\end{document}